\documentclass[12pt]{article}
\usepackage{amsmath}
\usepackage{graphicx}
\usepackage{enumerate}
\usepackage[round]{natbib}

\usepackage{url} 

\usepackage{multirow}
\usepackage{bm}
\usepackage{bbm}
\usepackage{dsfont}
\usepackage{amsfonts}
\usepackage{float}
\usepackage{comment} 

\graphicspath{ {./images/} }

\usepackage{xcolor}

\newcommand{\blind}{0}

\usepackage[margin = 0.9in]{geometry}

\newtheorem{theorem}{Theorem}

\begin{document}

\bibliographystyle{abbrvnat}

\def\spacingset#1{\renewcommand{\baselinestretch}%
{#1}\small\normalsize} \spacingset{1}


\if0\blind
{
  \title{\bf Individualized treatment regimens under correlated data with multiple outcomes
}

\author{Misha Dolmatov\thanks{
    The authors gratefully acknowledge the United Network for Organ Sharing for providing the motivating OPTN data, and the context for the work. The work is partially funded by grants R01DE031134 and R21DE031879 from the United States National Institutes of Health.}, Erica Moodie, David Stephens \hspace{.2cm}\\ 
    Department of Epidemiology, Biostatistics and Occupational Health, \\ McGill University, Montreal, QC, Canada\\ \\
    and \\ \\
    Dipankar Bandyopadhyay \\
    Department of Biostatistics, School of Public Health, \\ Virginia Commonwealth University, Richmond, VA, USA
    }

    \date{}
 \maketitle
} \fi

\if1\blind
{
  \bigskip
  \bigskip
  \bigskip
  \begin{center}
    {\LARGE\bf Individualized treatment regimens under correlated data with multiple outcomes}
\end{center}
  \medskip
} \fi

\vspace{-3mm} 
\begin{abstract}
Precision medicine involves developing individualized treatment regimes (ITRs) which allow for treatment decisions to be tailored to patient characteristics. Naturally, the identification of the optimal regime, that is, the rule which maximizes patient outcomes, is of interest. Several procedures for estimating optimal ITRs from observational data have been proposed; however, relatively few methods exist for estimating optimal ITRs in the presence of competing risks. Previous approaches either target one particular cause of failure, or rely on singly-robust estimators. We propose a novel doubly-robust regression-based method for estimating optimal ITRs which accounts for the uncertainty related to the unobserved cause of failure by averaging over all possible causes, or targeting the most likely cause. Our approach is straightforward to implement, and we demonstrate an extension to incorporate clustering, motivated by the question of for whom kidney transplantation with hepatitis C virus (HCV)-positive donors is safe, using data from the Organ Procurement and Transplantation Network. Our analysis suggests that a large portion of HCV-negative kidney recipients would see their overall survival unchanged if they were instead provided a kidney from an HCV-positive donor. The estimated treatment rules could be used to provide more efficient allocation of HCV-positive kidneys, increasing the donor pool. 
\end{abstract}

\noindent%
\vfill

\newpage
\spacingset{1.9} 


\section{Introduction}
The field of personalized medicine involves developing adaptive treatment rules which prescribe the most beneficial treatment based on available patient information. This framework is especially relevant when a given exposure has associated risks, which must be taken into account when assessing the overall benefit of treatment assignment. In such scenarios, patient covariate information can be used to guide decision-making and tailor treatment to an individual's personal level of risk. Covariate-tailored decision rules are referred to as individualized treatment regimes (ITRs). Naturally, the identification of the optimal regime, that is, the rule which maximizes patient outcomes, is of immense interest.

Our motivating question is focused on the problem of kidney transplantation with hepatitis C virus (HCV)-positive donors. Historically, kidney transplants from HCV-positive donors to HCV-negative recipients have been controversial due to concerns of viral transmission \citep{milford1991}. However, a recent study using data from the Organ Procurement and Transplant Network (OPTN) registry investigated the feasibility of using kidneys from HCV-positive donors in order to increase the donor pool \citep{gupta2017}, and showed that HCV-negative recipients with HCV-positive donors had lower graft and patient survival than their counterparts with HCV-negative donors. However, patients with a recipient-donor mismatch in HCV status still had superior survival outcomes when compared to those who remained on the waitlist. These results point towards mismatched transplants being a viable alternative for certain patient subgroups. In other words, if one could identify the patient subgroups that are less at risk of poor outcomes due to the HCV status of their transplant donor, this would yield an easy way to increase the number of available donors while limiting adverse events. We revisit the OPTN dataset, approaching this problem from the ITR perspective in order to identify additional tailoring variables which may permit the use of HCV+ kidneys for safe transplantations. 

Estimation of the optimal regime from observational data requires careful consideration of the relationship between the treatment assignment mechanism, patient covariate information, and the study design. Several regression-based approaches for estimating ITRs from observational data have been proposed, including Q-learning \citep{murphy2003}, G-estimation \citep{Robins2004}, and dynamic weighted ordinary least squares \citep[dWOLS;][]{WallacedWOLS}. Variants of these methods that allow for time-to-event (``survival'') outcomes subject to right-censoring have also been suggested. For example, dynamic weighted survival modelling \citep[DWSurv;][]{Simoneau2020} was developed as an extension to the doubly-robust weighted linear regression approach of dWOLS. The method aims to estimate the optimal regime via weighted accelerated failure time \citep[AFT;][]{orbe2006alternative} models at each stage of treatment with inverse probability of being censored (IPC) weights, combined with balancing weights (such as overlap or inverse probability of treatment weights) to account for non-random censoring and non-randomized treatment assignment.

An important consideration when dealing with survival data is that of competing risks, i.e., events that prevent the occurrence of other types of events \citep{prentice1978}. For example, in kidney transplantation, if treatment recommendations were made to maximize the time to graft rejection (post transplantation), deaths due to a cause unrelated to graft failure would constitute a competing risk. Despite being of vast clinical importance, research on the topic has been relatively limited in the dynamic treatment regime (DTR) literature. Most of the methods developed to identify competing risk DTRs have revolved around providing novel ways of estimating the Fine and Gray \citep{Fine1999} cumulative incidence function (CIF) for the event type of interest in multi-stage designs \citep{Yavuz2016, Chen2018}. For instance, \cite{Morzywołek2022} estimates the CIFs associated with a fixed regime via dynamic-regime marginal structural models. In contrast, a recent work \citep{Choi2019} has investigated the possibility of using AFT models for competing risk problems, allowing for treatment effects to be directly interpreted in terms of the survival time.

The contributions discussed above typically consider each event type separately for DTR estimation. However, since the cause of failure of each subject is unknown at treatment assignment, optimal decision-making requires consideration of all competing risks in order to maximize overall event-free survival. One possible solution, based on Q-learning \citep{clifton2020q}, estimates a set-valued DTR that results in a set of recommended treatments that are not unfavourable according to any of the competing outcomes \citep{laber2014}. As Q-learning is a singly-robust procedure, this approach is inherently more sensitive to model misspecification. Finally, none of the methods proposed thus far allow for clustered data (subjects within centers), which are of particular interest in precision medicine, with large multi-center studies often being required for sufficient power to investigate treatment effect heterogeneity.


This article extends the weighted AFT methodology of DWSurv to account for competing risks and clustering. Our method, based on a doubly-robust implementation of generalized estimating equations (GEE), permits consistent estimation of parameters of interest while allowing misspecification of a subset of nuisance models employed to address confounding, loss to follow up, or the median survival time itself. The estimated optimal treatment rule is expressed as a function of the optimal ITRs for each competing cause of failure, in order to account for the uncertainty regarding the true cause of failure at treatment allocation. Our flexible approach allows for different functions to be considered and compared based on the resulting treatment rule and its ability to maximize the overall survival time. The proposed extensions provide straightforward, regression-based methods for determining optimal ITRs in the presence of competing risks.

The rest of the paper is organized as follows. Starting with definitions and assumptions, Section~\ref{sec:methodology} presents the development of the optimal ITR, related estimation and inference, and regime evaluation. Section~\ref{sec:sim} explores numerical characteristics and behaviour of the proposed ITRs in a number of finite sample simulation settings. The proposed methodology is illustrated via application to the motivating OPTN kidney transplant allocation data in Section~\ref{sec:application}. Finally, Section~\ref{sec:conclusion} concludes, with a discussion. Additional materials, consisting of plots and tables from simulation studies and real data illustration, and some theoretical details are relegated to Appendices A-C.

\section{Methodology} \label{sec:methodology}

\subsection{Definitions and assumptions}
\label{subsec:defs}
To each subject, we associate the vector of observable quantities $(\bm{X}, A, \Delta, C, K, R, T)$. Within the competing risk framework, let $K \in \{1, \ldots, \kappa\}$ be the cause of failure indicator, and $T$, the associated failure time. Without loss of generality, we assume that a longer time to event (failure time) is more clinically desirable. Let $A \in \mathcal{A}$ represent the treatment assignment and $\bm{X} \in \mathcal{X}$ the vector of measured pre-treatment covariates for each subject. Denote by $\Delta$, the event indicator and by $C$, the censoring time, where $\Delta = 1$ if an event of any type was observed and $0$ otherwise. Note that neither $K$ nor $T$ are observed when $\Delta = 0$.  Finally, let $R \in \{1, \ldots, r\}$ denote the cluster membership indicator. Although not immediately relevant for defining the optimal treatment rule, accounting for clustering in the outcome variable is important for efficient parameter estimation. This will be discussed further in Subsections \ref{subsec:aft} and \ref{subsec:estimation}. The observed data is the collection of random variates $\{(\bm{X}_i, A_i, \Delta_i, (1-\Delta_i)C_i, \Delta_iK_i, R_i, \Delta_i T_i)\}_{i=1}^n$. We define $T(a,k)$ as the counterfactual survival time for an individual if, possibly contrary to the fact, they had been assigned treatment $a$ and failed from cause $k$. An ITR, in our case, is a function $d : \mathcal{X} \to \mathcal{A}$ that maps a subject's covariate information to the set of possible treatments. Similarly, we can define $T(d,k)$ as the potential outcome of an individual with cause of failure $k$ and treatment assigned according to $d$. For causal identification and estimation of these potential outcomes, we make the following assumptions:

\begin{enumerate}
    \item Stable unit treatment value (SUTVA): The failure time of a subject is independent of other patients' treatment assignment \citep{rubin1980}.
    \item Consistency: Observed and counterfactual outcomes agree for $A = a$ and $K = k$, i.e., $ T = \sum_{a \in \mathcal{A}, k =1,\ldots, \kappa} \mathds{1}_{a}(A)\mathds{1}_{k}(K)T(a,k).$
    \item No unmeasured confounders (NUC) \citep{gill2001causal}: $T(a,k) \perp (A,K) | \bm{X} \\ \forall a \in \mathcal{A}, \;  \forall k = 1, \ldots, \kappa. $
    \item Coarsening at random \citep{gillLaanRobins1997}: $T(a,k) \perp \Delta |\bm{X} \; \forall a \in \mathcal{A}, \; \forall k = 1, \ldots, \kappa.$
    \item Conditional independence of failure type and joint distribution of treatment and censoring mechanisms: $K \perp (A, \Delta)|\bm{X}.$
\end{enumerate}

The fundamental problem of ITR estimation in the competing risk setting is that the true future cause of failure for any patient is unknown at treatment assignment. As a result, optimal decision-making must account for this uncertainty as well as treatment effect heterogeneity. If the true cause of failure was known at treatment assignment, i.e., $K$ was among the vector of measured covariates for each subject, the optimal ITR could be directly specified as
\begin{equation}\label{eq:oracle_dtr}
    d^{\text{opt}}_{\text{o}}(\bm{X}, K) = \arg \max_{d} \mathbb{E}[f(T(d,K)) - \zeta(d,\bm{X})] 
\end{equation}
for some fixed monotonically increasing function $f: \mathbb{R}_+\to \mathbb{R}$ and some known cost function $\zeta : \mathcal{A} \times\mathcal{X} \to \mathbb{R}_+$, which represents some measure of financial or other cost associated to the allocation of treatment on the transformed survival scale. We will refer to this regime as the \textit{oracle} regime. The cost function can also be used to specify a minimal effectiveness threshold, with treatment only being assigned when the treatment effect is larger than this threshold. This interpretation is made more concrete in Subsection \ref{subsec:aft}. Alternatively, in resource-limited settings, $\zeta$ can be specified as a data-dependent constant which determines the proportion of the study population that will receive treatment \citep{Luedtke2016}. 

A natural definition of the optimal ITR for unknown $K$ is one that marginalizes over the distribution of the cause of failure $K$ in equation (\ref{eq:oracle_dtr}):
\begin{align*}
    d^{\text{opt}}_{\text{w}}(\bm{X}) &= \arg \max_{d} \mathbb{E}\Bigl[f(T(d,K)) - \zeta(d,\bm{X})\Bigr] ,\\
    &= \arg \max_{d} \mathbb{E}\Bigl[\mathbb{E}[f(T(d,K))| \bm{X}] - \zeta(d, \bm{X})\Bigr],\\
    &=  \arg \max_{d} \mathbb{E}\left[ \sum_{k=1}^\kappa P(K = k| \bm{X}) \mathbb{E}[f(T(d,k))| \bm{X}]- \zeta(d, \bm{X})  \right]. 
\end{align*}

%

In the DTR literature, the optimal treatment rule is often expressed in terms of causal contrasts with respect to some reference treatment $A = 0$, leading to the following definition:
\begin{align*}
    \gamma_k(\bm{x}, a) = \mathbb{E}[f(T(a,k)) - f(T(0,k)) | \bm{X} = \bm{x}], \quad k = 1, \ldots, \kappa \text{ and } \bm{x} \in \mathcal{X}. 
\end{align*}
This quantity is called the \textit{blip} function and represents a causal contrast (often a difference) in outcomes --- here in the transformed survival times --- of an individual with covariates $\bm{X} = \bm{x}$ when comparing treatment $A = a$ with the reference treatment \citep{Robins2004}. The expected counterfactual survival time can then be written in terms of the blip and a second component which does not depend on treatment:
\begin{align*}
    \mathbb{E}[f(T(a,k))|\bm{X}=\bm{x}] &= \mathbb{E}[f(T(0,k)) | \bm{X}=\bm{x}]+ \mathbb{E}[f(T(a,k)) - f(T(0,k)) | \bm{X}=\bm{x}],\\
    &= m_k(\bm{x}) + \gamma_k(\bm{x},a), \quad k = 1, \ldots, \kappa.
\end{align*}
The first term $m_k(\bm{x})$ is designated the \textit{treatment-free} model as it only depends on the covariates $\bm{X}$ and the chosen reference treatment, but not $a$. Thus, the optimal ITR can be expressed in terms of the blip functions for each failure type as
\begin{align}
    d^{\text{opt}}_{\text{w}}(\bm{X}) &= \arg \max_{d} \mathbb{E}\left[ \sum_{k=1}^\kappa \varphi_k(\bm{X}) \gamma_k(\bm{X}, d)  - \zeta(d, \bm{X})\right],
\end{align}
where, $\varphi_k(\bm{X}) = P(K =k|\bm{X})$. We will refer to this ITR as the \textit{weighted} rule. As a result, determining the optimal regime relies on the identification of the blip functions for all $K$ failure types. Under assumptions 2 and 3, the blip functions can be written in the following form:
\begin{align*}
    \gamma_k(\bm{x}, a) &= \mathbb{E}[f(T(a,k)) - f(T(0,k)) | \bm{X} = \bm{x}] ,\\
    &= \mathbb{E}[f(T)|\bm{X}=\bm{x}, A = a,K=k] - \mathbb{E}[f(T)|\bm{X}=\bm{x}, A=0,K=k] \quad k = 1,\ldots, \kappa.
\end{align*}
Given that that this contrast is expressed as the difference in conditional means, a natural approach would be to posit regression models for the survival times associated to each failure type, and estimate the blip directly. However, outcome regression-based methods are known to be sensitive to model misspecification; this considerable weakness will be addressed in Subection \ref{subsec:estimation} when discussing doubly-robust estimation.

Equation (\ref{eq:oracle_dtr}) can be modified to obtain other definitions of an optimal ITR. For instance, we can consider a rule which optimally treats the most probable cause: 
\begin{align*}
d^{\text{opt}}_{\text{g}}(\bm{X}) &= \arg\max_d  \mathbb{E}[f(T(d, \arg\max_{k} \varphi_k(\bm{X}))) - \zeta(d,\bm{X})],\\
 &-\arg\max_d  \mathbb{E}[\gamma_{k^*}(\bm{X},d) - \zeta(d,\bm{X})], \quad k^*=\arg\max_{k} \varphi_k(\bm{X}),
\end{align*}
which we designate the \textit{greedy} rule. Alternatively, the functions $\varphi_k(\bm{x})$ can be replaced by fixed constants which reflect the beliefs of the analyst concerning the relative importance of each cause in the decision making process.

\subsection{Accelerated failure time specification} \label{subsec:aft}

A common choice for $f$ is $f(t) = \log t$, resulting in an AFT model associated to each cause. For the remaining sections, we assume binary treatment, i.e., $\mathcal{A} = \{0,1\}$. Suppose the data are comprised of $r$ clusters, indexed by $i = 1, \ldots, r$, each of size $r_i$. Assume that the number and size of these clusters is fixed. For each cluster $i = 1, \ldots, r$, we consider the following specification for the mean of the response vector $\mathbf{T_i}$: 
\begin{align*}
      \log \mathbf{T_i}|\mathbf{K_i}=\mathbf{k_i} &\sim \mathbf{X}_{\bm{\beta_k}} \bm{\beta_k} + \mathbf{A}  \mathbf{X}_{\bm{\psi_k}}\bm{\psi_k} + \boldsymbol{\varepsilon_i},
\end{align*}
where, $ \mathbb{E}[\boldsymbol{\varepsilon_i}] = \boldsymbol{0}$ and $\text{var}[\boldsymbol{\varepsilon_i}] = \boldsymbol{V_i}$. The matrix $\mathbf{X}_{\bm{\psi_k}}$ contains all components of $\bm{X}$ that modify the effect of treatment for the $k$-th failure type, augmented with a leading column of $1$s in order to account for the main effect of treatment. The matrix $\mathbf{X}_{\bm{\beta_k}}$ contains the same features as $\mathbf{X}_{\bm{\psi_k}}$, including the leading column of $1$s, along with relevant confounders and other variables predictive of the outcome in the absence of treatment. Using the same identification arguments as in the previous subsection, we can connect this parametrization to the causal quantities defined above:
\begin{align*}
    \mathbb{E}[\log T |\bm{X} = \bm{x}, A = a,K=k] &= m_k(\bm{x}_{\bm{\beta_k}}; \bm{\beta_k}) + \gamma_k(\bm{x}_{\bm{\psi_k}}, a; \bm{\psi_k}),\\
    &= \bm{x}_{\bm{\beta_k}}\bm{\beta_k} + a\bm{x}_{\bm{\psi_k}}\bm{\psi_k}, \quad k = 1, \ldots, K.
\end{align*}
This allows us to directly write the regime $d^{\text{opt}}_{\text{w}}$ in terms of the parameters $\bm{\psi_k}$:
\begin{align*}
     d^{\text{opt}}_{\text{w}}(\bm{X}) &= \arg \max_{d} \mathbb{E}\left[ d(\bm{X})\sum_{k=1}^K \varphi_k(\bm{X})\bm{X}_{\bm{\psi_k}}\bm{\psi_k} - \zeta(\bm{X},d)\right]
\end{align*}
Since the treatment is binary, we can write the cost function as $\zeta(\bm{X},d) = \zeta_0(\bm{X}) + d(\bm{X})\zeta_1(\bm{X})$. As a result, $d^{\text{opt}}_{\text{w}}$ can be directly ascertained as
\begin{align*}
    d_{\text{w}}(\bm{x}) =\mathds{1}\left(\sum_{k=1}^\kappa \varphi_k(\bm{x})\bm{x}_{\bm{\psi_k}} \bm{\psi_k} > \zeta_1(\bm{x})\right),\quad \bm{x} \in \mathcal{X}.
\end{align*}
The weighted rule has an intuitive interpretation: it prescribes treatment as long as the treatment has an overall effect larger than some minimal threshold $\zeta_1(\bm{x})$, where this global effect is a convex combination of the effects of treatment under different failure types, with weights corresponding to the probability of the occurrence of each cause of failure. As a result, if treatment dramatically decreases survival times for any given failure type, the rule might possibly recommend withholding treatment, even if failures of that type are rare and treatment has a protective effect with respect to the other failure types; in that case, the risk outweighs the benefits. 

Following the derivations above, we can write the oracle and greedy regimes in terms of the blip parameters:
\begin{align*}
    d_{\text{o}}(\bm{x}, k) &= \mathds{1}\left(\bm{x}_{\bm{\psi}_{k}}\bm{\psi}_{k} > \zeta_1(\bm{x})\right),\\
    d_{\text{g}}(\bm{x}) &=\mathds{1}\left(\bm{x}_{\bm{\psi}_{k^*}}\bm{\psi}_{k^*} > \zeta_1(\bm{x})\right),\quad  k^*=\arg\max_{k} \varphi_k(\bm{x}), \quad \bm{x} \in \mathcal{X}.
\end{align*}
All treatment rules $d$ defined above can be written as $d(\bm{x}) = \mathds{1}(\mathcal{B}_{d}(\bm{x}) > \zeta_1(\bm{x}))$
for some function $\mathcal{B}_{d}: \mathcal{X}\to \mathbb{R}$, which we term the \textit{benefit} of the ITR $d$. In particular,
\begin{align*}
    \mathcal{B}_{d_{\text{w}}}(\bm{x}) &= \sum_{k=1}^\kappa \varphi_k(\bm{x})\bm{x}_{\bm{\psi_k}} \bm{\psi_k},\\
    \mathcal{B}_{d_{\text{g}}}(\bm{x}) &=\bm{x}_{\bm{\psi}_{k^*}}\bm{\psi}_{k^*}, \quad k^*=\arg\max_{k} \varphi_k(\bm{x}),\\
    \mathcal{B}_{d_{\text{o}}}(\bm{x}, k)&= \bm{x}_{\bm{\psi}_{k}}\bm{\psi}_{k}.
\end{align*}
The benefit for the oracle regime depends on the true cause of failure, $k$.

\subsection{Estimation and inference}
\label{subsec:estimation}
In Subsection~\ref{subsec:aft}, we derived the form of the optimal treatment rule in terms of the blip parameters. Consequently, determining the optimal ITR relies on consistent estimation of these parameters. For this purpose, we consider DWSurv, which provides doubly-robust semiparametric estimation of blip parameters in AFT models with weights accounting for non-randomized treatment assignment and covariate-dependent right-censoring \citep{Simoneau2020}. The weighted GEE approach used in DWSurv can easily be modified to allow for non-independence of observations by specifying a working covariance matrix for each cluster \citep{zeger1986}. 

We propose estimating all $\kappa$ sets of blip parameters by applying DWSurv to the observations of each of the failure types separately. The estimation procedure for an arbitrary treatment rule $d(\bm{x}; \bm{\psi})$ is summarized by the following algorithm:

\begin{enumerate}
    \item Posit models for the treatment, censoring, and cause of failure mechanisms: $P(A = 1|\bm{X}=\bm{x}; \bm{\alpha})$, $P(\Delta =1|A=a,\bm{X}=\bm{x}; \bm{\xi})$, and $P(K = k| \bm{X}=\bm{x}; \bm{\eta})$ respectively. Obtain estimates $\hat{\bm{\alpha}}, \hat{\bm{\xi}}$, and $\hat{\bm{\eta}}$ from the observed data. 
    \item For failure types $k = 1, \ldots, \kappa$,
    \begin{enumerate}
        \item Propose a linear model for the outcome of the form $\mu = \mathbb{E}[\log T|\bm{X}=\bm{x}, A=a, K = k; \bm{\beta_k}, \bm{\psi_k}] = \bm{x}_{\bm{\beta_k}}\bm{\beta_k} + a\bm{x}_{\bm{\psi_k}}\bm{\psi_k}$, and propose a working covariance matrix $\mathbf{V}_i(\bm{\lambda})$ for all clusters $i = 1, \ldots, r$.
        \item Choose a weight function $w_k(\delta, a, \bm{x}; \alpha)$ and compute the associated weights $\hat{w}_k$ with the estimates $\hat{\bm{\alpha}}$ and $\hat{\bm{\eta}}$. See below for details.
        \item Given initial estimate $\hat{\bm{\lambda}}^{(0)}$, set $t = 0$ and alternate between the following two steps until convergence is achieved:
        \begin{enumerate}
            \item[i.] Obtain estimate of $\hat{\bm{\theta}}^{(t+1)}= (\hat{\bm{\beta_k}}^{(t+1)}, \hat{\bm{\psi_k}}^{(t+1)})$ by solving the weighted estimating equation
            \begin{align}\label{eq:weighted_gee}
                \sum_{i=1}^r \mathbf{D}_i^\top [\mathbf{V}_i (\hat{\bm{\lambda}}^{(t)})]^{-1}\; \mathbf{W_i}\Bigl(\log (\mathbf{T}_i) - \bm{\mu}_i\Bigr)=0,
            \end{align}
            
            where $ \mathbf{D}_i = \frac{\partial \bm{\mu}_i}{\partial \bm{\theta} } = [\mathbf{X}^{(i)}_{\bm{\beta_k}} \;\; \mathbf{A}^{(i)}  \mathbf{X}^{(i)}_{\bm{\psi_k}}], \quad
                \mathbf{W}_i = \text{diag}\Bigl(\delta_{ij} \hat{w}_{ijk}\Bigr)_{j = 1, \ldots, r_i}.$
            \item[ii.] Estimate $\hat{\bm{\lambda}}^{(t+1)}$ via method of moments based on the estimated residuals \\  $\hat{\bm{e}}_{i}(\hat{\bm{\beta_k}}^{(t+1)}, \hat{\bm{\psi_k}}^{(t+1)}) = \log (\mathbf{T}_i) - \hat{\bm{\mu}}_i^{(t+1)}$.
        \end{enumerate}
    \end{enumerate}
    \item Obtain the estimated optimal treatment rule $d(\bm{x};\bm{\hat{\psi}}, \hat{\bm{\eta}})$ from the final estimates of the blip parameters and the parameters of the failure type model.
\end{enumerate}

The weights are chosen to satisfy the balance condition stated in the following theorem (proof in Appendix C). 
\begin{theorem}
    Under assumptions $1$-$5$ (see Subsection \ref{subsec:defs}), solving the weighted GEE (\ref{eq:weighted_gee}) yields consistent estimators for $\{\bm{\psi}_k\}_{k=1}^\kappa$ if the weights satisfy the following balancing property: 
    \begin{align*}
        [1- c(0, \bm{x})][1- b(\bm{x})]w_k(0,0,\bm{x}) &= c(0,\bm{x})[1-b(\bm{x})]w_k(1,0,\bm{x})\\
    &=  [1- c(1, \bm{x})]b(\bm{x})w_k(1,0,\bm{x})\\
    &= c(1,\bm{x})b(\bm{x})w_k(1,1,\bm{x}), \quad \text{ for }k = 1, \ldots, \kappa, 
    \end{align*}
    where $b(\bm{x}) =  P(A =1 | \bm{X}=\bm{x})$ and $c(a, \bm{x}) = P(\Delta =1 |A =a, \bm{X}=\bm{x})$.
\end{theorem}
 For example, weights of the form
\begin{equation}\label{eq:overlap} 
    w(\delta, a, x) = \frac{|a - P(A =1 |\bm{X}=\bm{x})| }{P(\Delta =\delta |A=a,\bm{X}=\bm{x})}
\end{equation}
satisfy the above equation \citep{Simoneau2020}.

Estimating the blip parameters $\bm{\psi_k}$ via DWSurv offers the advantage of double-robustness. For each of the failure types $k = 1, \ldots, \kappa$, under correct specification of the blip function $\gamma_k$, it suffices to correctly specify either the treatment-free model or the weights in order to obtain consistent estimation of $\bm{\psi_k}$. We note that correctly specifying the weights requires correct models for both the treatment and censoring mechanisms. Furthermore, under standard regularity conditions for GEEs, the procedure outlined above is robust to misspecification of the working correlation structure: as long as either the outcome model or the weighting models are correctly specified, we have consistent estimation of the blip parameters, albeit with a possible loss of efficiency \citep{zeger1986}.

If the nuisance models are estimated by maximum likelihood, the consistency and asymptotic normality of the blip parameter estimators follows directly from standard results on joint estimating equations \citep{Vaart_1998}. Otherwise, similar conclusions hold under additional assumptions regarding the convergence of nuisance parameter estimators \citep{yuan2000}. The asymptotic variance of  $\hat{\bm{\psi}}$ can be obtained by considering a Taylor expansion of the estimating equation presented in step c(i) about the limiting distributions of the nuisance parameters \citep{Simoneau2020}. However, due to the large number of nuisance parameters, deriving an explicit formula is very cumbersome and consequently, of limited practical use. The nonparametric cluster bootstrap offers a conceptually simpler alternative to variance estimation at the cost of a larger computational burden \citep{DavisonHinkley1997}.  Alternatively, the sandwich errors that ignore the estimation of the weights may often provide reasonable performance \citep{davidian2004}.

\subsection{Regime evaluation}
\label{subsec:metrics}

Although the treatment rules presented above target different criteria, a particular ITR might be preferable in a given situation based on overall in-sample performance. For a given regime $d$, we consider two main measures of performance, namely the proportion of optimal treatment (POT) and the value. These are defined as follows: 
\begin{align*}
    \text{POT}(d) &= \mathbb{E}[\mathds{1}(d(\bm{X}) = d_{\text{o}}(\bm{X}, K))],\\
    \text{V}(d) &= \mathbb{E}\left[\log T(d,K)\right].
\end{align*}
The POT for a decision rule $d$ measures the proportion of individuals for whom the treatment allocated under $d$ is equal to the treatment prescribed by the oracle regime. The value of $d$ is the average survival time of a subject within the study population whose treatment is assigned according to $d$. The ideal ITR targets the specified criteria while also maintaining reasonably high POT and value. The following two unbiased estimating equations in $p$ and $\mu$ yields estimators of the POT and value of an ITR $d$, respectively:
\begin{align*}
    \frac{1}{n} \sum_{i=1}^n \frac{\Delta_i}{P(\Delta = 1 | \bm{X}_i, A_i)} \Bigl(\mathds{1}(d(\bm{X}_i) = d_{\text{o}}(\bm{X}_i, K_i)) - p \Bigr) &= 0,\\
    \frac{1}{n} \sum_{i=1}^n \frac{\Delta_i}{P(\Delta = 1 |\bm{X}_i, A_i)} \Bigl( \log T_i + (d-A_i) \bm{X}_{\psi_{K_i}} \bm{\psi}_{K_i} - \mu \Bigr) &= 0.
\end{align*}

Proofs of unbiasedness can be found in Appendix C. Since both the true censoring probabilities and the true values of the blip parameters (and thus, the oracle regime) are unknown, they must be estimated from the data. Estimates are readily available following the estimation procedure detailed in Subsection~\ref{subsec:estimation}. Under correct specification of the censoring model and of the blip functions for each cause, as well as standard regularity conditions, the POT and value estimators are consistent and asymptotically normal. In practice, the estimating equations above can be solved by fitting intercept-only linear models to the outcomes $\mathds{1}(d(\bm{X}_i) = \hat{d}_{\text{o}}(\bm{X}_i, K_i))$ and $\log T_i + (d-A_i) \bm{X}_i \hat{\bm{\psi}}_{K_i}$, with estimated inverse probability of censoring weights $\hat{w}_i = \Delta_i /\hat{P}(\Delta = 1 | \bm{X}_i, A_i)$.  The following Subsection highlights the connection between ITR estimation with competing risks and ITR estimation for outcomes generated by a finite mixture model.

\subsection{Connection to finite mixture models ITRs}
\label{subsec:mixture}

We consider a binary treatment, continuous outcome setting without any censoring. Let ($Y, \bm{X}, A$) denote the triplet of observed variables, with $Y$ the outcome variable. Suppose $Y$ is generated from a finite mixture model with $\kappa$ components, each with the same linear decomposition as in Subsection \ref{subsec:aft}. In particular, for $k = 1, \ldots, \kappa$, we have that the conditional distribution of $Y$ given $K = k$ is
\begin{align*}
    Y | K = k \sim \bm{X}_{\beta_k} \bm{\beta}_k + A \bm{X}_{\psi_k} \bm{\psi_k} + \bm{\varepsilon},
\end{align*}
where, $K$ denotes the latent mixture component indicator, and $\mathbb{E}[\bm{\varepsilon}] = 0$.

Let $Y(d, k)$ denote the counterfactual for a subject with treatment assigned via regime $d$ and whose outcome was generated according to the $k$-th mixture component. Assuming the equivalent versions of Assumptions 1, 2, 3, and 5, we can derive ITRs which take into account the uncertainty associated with the unknown mixture membership indicator $K$:
\begin{align*}
    d_w &= \arg \max_{d} \mathbb{E}[Y(d,K)] = \mathds{1}\left(\sum_{k=1}^\kappa \varphi_k(\bm{x})\bm{x}_{\bm{\psi_k}} \bm{\psi_k} > 0\right),\\
    d_g &= \arg\max_d \mathbb{E}[Y(d, \arg\max_{k} \varphi_k(\bm{X}))] =\mathds{1}\left(\bm{x}_{\bm{\psi}_{k^*}}\bm{\psi}_{k^*} > 0\right),\quad  k^*=\arg\max_{k} \varphi_k(\bm{x}),
\end{align*}
where, all quantities are defined as before. These ITRs have the same interpretations as those introduced in Subsection \ref{subsec:defs}; the weighted regime averages over all mixture components while the greedy regime targets the most likely component of the mixture distribution. However, in the finite mixture model case, $K$ is not observed and thus blips cannot be estimated by considering observations from each mixture component separately as in the competing risk setting. The expectation-maximization (EM) algorithm, Bayesian data augmentation, or other standard estimation methods for finite mixture models can instead be used to estimate the blips under missingness of the mixture membership indicator \citep{mclachlan2019}. However, these approaches do not possess the same robustness properties as the method detailed above.

\section{Simulations}
\label{sec:sim}

We conducted several simulation studies to evaluate the finite-sample properties of the proposed estimation procedure and to compare the different treatment rules presented in Subsection~\ref{subsec:aft}. As in Subsection~\ref{subsec:estimation}, we assume binary treatment; without loss of generality, we also assume $\zeta(\bm{x}) = 0$, i.e., no costs associated with treatment allocation.  

In the simulations, blip parameters were estimated from a training set, with the resulting ITRs then  evaluated on a fixed independent uncensored test set under the same data-generating mechanism. Default samples sizes of $n_{\text{train}} = 1000$  and $n_{\text{test}} = 10,000$ were used for the training and testing sets, respectively, unless specified otherwise. The variability of the estimating procedure was assessed by repeating parameter estimation over ($n_{\text{rep}} = 1000$) replicate training sets. Estimated standard errors (SE) and biases were reported for all blip parameter estimators. The various ITRs defined above were compared based on their proportion of optimal treatment and their value, as defined in Subsection \ref{subsec:metrics}. In this case, the expectation is taken with respect to the test set over all replications. Since the test set is uncensored, the POT and value were estimated via sample averages.

For the sake of comparison, we also define the \textit{uniform} regime, which allocates treatment by sampling uniformly on $\mathcal{A} = \{0,1\}$: $d_{\text{u}}(\bm{x}) \sim \text{Bernoulli}(0.5).$
Across all simulations scenarios, the estimated weighted and greedy ITRs were compared to the oracle and uniform ITRs, which served as benchmark regimes. The oracle regime was defined in terms of the true data-generating values of the blip parameters in order to provide an upper bound on overall performance.

\subsection{Data generating mechanism}
\label{subsec:dgm}
Let $\text{expit}(v) = \exp(v)/(1 + \exp(v))$ for $v\in \mathbb{R}$. To allow for dependence between observations, survival times for each individual were generated via an AFT model with a random intercept, corresponding to an exchangeable within-cluster correlation structure on the log-survival scale. For cluster $i = 1,\ldots, r$ and subject $j = 1, \ldots, r_i$, we have the following model specification:
\begin{align*}
    X_{ij1} &\sim \mathcal{N}(0,1), \quad A_{ij} \sim \text{Bernoulli}\Bigl(\text{expit}(0.5 + X_{ij1} + X_{ij2})\Bigr),\\
    X_{ij2} &\sim \mathcal{N}(0,4), \quad \Delta_{ij} \sim \text{Bernoulli}\Bigl(\text{expit}(\delta_0 - X_{ij1} -0.3 X_{ij2})\Bigr),\\
     K_{ij} &\sim \text{Bernoulli}\Bigl(\text{expit}(0.5 + X_{i1})\Bigr) + 1,\\
     U_i&\sim \mathcal{N}(0,\tau^2), \quad \varepsilon_{ij1},\varepsilon_{ij2}  \sim \mathcal{N}(0, \sigma^2),\\
     T_{ij}|K_{ij} = 1 &\sim \exp\Bigl(1 + 0.5 X_{ij1} - 0.3X_{ij2} + U_i + A_{ij}(\psi_{11} + \psi_{12}X_{i1}) + \varepsilon_{ij1}\Bigr),\\
     T_{ij}|K_{ij} = 2 &\sim \exp\Bigl(2 -0.1 X_{ij1} + 0.2X_{ij2} + U_i + A_{ij}(\psi_{21} +\psi_{22} X_{i1}) + \varepsilon_{ij2}\Bigr).
\end{align*}
Cluster membership was determined by drawing from an independent uniform distribution on $\{1, \ldots, r\}$.  The following default values were used for all simulation settings, unless specified otherwise: $r = 50, \; \delta_0 = 1.73,\;   \tau^2 + \sigma^2 = 0.5$ and $\text{ICC} = \tau^2/\sigma^2 + \tau^2 = 0.5$.

This data generating mechanism yields a censoring rate of approximately $25\%$, with $40\%$ of individuals failing from cause $1$.  Estimation for the weighting models and the cause of failure model was accomplished using logistic regression, while estimation for the AFT-GEE model was done via the \texttt{geem} package in R. Weights as in (\ref{eq:overlap}) were used for fitting the weighted GEE. Unless otherwise stated, the exchangeable structure was used for the working covariance matrix associated to each cluster. Both weighting models and outcome models were correctly specified for all simulation scenarios except for those in Subsection~\ref{subsec:param_sim}. The models for the cause of failure $K$ were correctly specified for all simulations. 
Additional simulations that explore the sensitivity of the proposed estimation procedure to variations in the data generating process and other modelling choices can be found in Appendix A.

\subsection{Parameter estimation}
\label{subsec:param_sim}
The first set of simulations showcases the consistency and double robustness of the estimation procedure outlined in Subsection \ref{subsec:estimation}. For this purpose, we consider two baseline settings, one with a moderate sample size, and one with a large sample size. For both settings, we conducted parameter estimation under four different model specifications:
\begin{enumerate}
    \item[(i)]  All models misspecified:
    \begin{align*}
        &P(A_{ij} = 1) = \text{logit} (\alpha_1 + \alpha_2 X_{ij1}), \quad P(\Delta_{ij} = 1) = \text{logit} (\alpha^*_1 + \alpha_2^* X_{ij1}),\\
        &\mathbb{E}[\log T|\bm{X}, A=a, K = k] = \beta_{k1} + \beta_{k2} X_{ij1} + a(\psi_{k1} + \psi_{k2} X_{i1}), \quad k = 1,2.
    \end{align*}
    \item[(ii)] Outcome model correct, weighting models incorrect:
    \begin{align*}
        &P(A_{ij} = 1) = \text{logit} (\alpha_1 + \alpha_2 X_{ij1}), \quad P(\Delta_{ij} = 1) = \text{logit} (\alpha^*_1 + \alpha_2^* X_{ij1}),\\
        &\mathbb{E}[\log T|\bm{X}, A=a, K = k] = \beta_{k1} + \beta_{k2} X_{ij1} + \beta_{k3} X_{ij2} +  a(\psi_{k1} + \psi_{k2} X_{i1}), \quad k = 1,2.
    \end{align*}
    \item[(iii)] Weighting models correct, outcome model incorrect:
    \begin{align*}
        &P(A_{ij} = 1) = \text{logit} (\alpha_1 + \alpha_2 X_{ij1} + \alpha_3 X_{ij2}), \; P(\Delta_{ij} = 1) = \text{logit} (\alpha^*_1 + \alpha_2^* X_{ij1} + \alpha_3^* X_{ij2}),\\
        &\mathbb{E}[\log T|\bm{X}, A=a, K = k] = \beta_{k1} + \beta_{k2} X_{ij1} + a(\psi_{k1} + \psi_{k2} X_{i1}), \quad k = 1,2.
    \end{align*}
    \item[(iv)] All models correctly specified:
    \begin{align*}
        &P(A_{ij} = 1) = \text{logit} (\alpha_1 + \alpha_2 X_{ij1} + \alpha_3 X_{ij2}), \; P(\Delta_{ij} = 1) = \text{logit} (\alpha^*_1 + \alpha_2^* X_{ij1} + \alpha_3^* X_{ij2}),\\
        &\mathbb{E}[\log T|\bm{X}, A=a, K = k] = \beta_{k1} + \beta_{k2} X_{ij1} + \beta_{k3} X_{ij2} +  a(\psi_{k1} + \psi_{k2} X_{i1}), \quad k = 1,2.
    \end{align*}
\end{enumerate}
Data-generating parameter values for setting 1 were (a) $(\psi_{11}, \psi_{12}) = (0.2, -0.2)$, $  (\psi_{21}, \psi_{22}) = (0.2, 0.2)$, with $n_{\text{train}} =1000$ and $r= 50$, and (b) $\bm{\psi}$ as before, with $n_{\text{train}} =5000$ and $r=250$.

\begin{figure}[H]
    \centering
    \includegraphics[scale = 0.53]{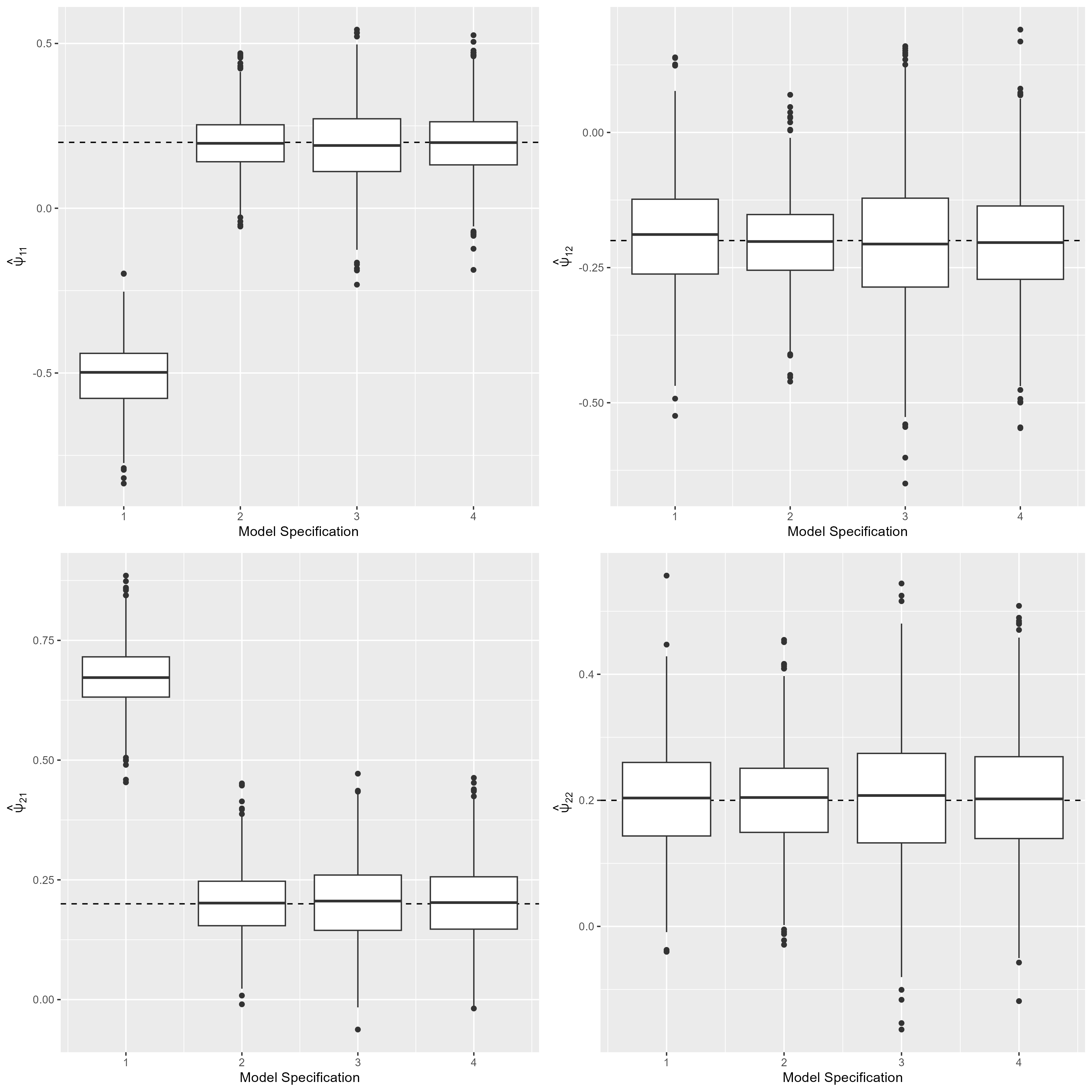}
    \caption{Empirical distribution of blip parameter estimates obtained via the weighted AFT-GEE model with sample size $n_{\text{train}} = 1000$ in setting 1(a) across 4 scenarios: (i) outcome and weighting models misspecified, (ii) outcome model correctly specified, (iii) weighting models correctly specified, (iv) all models correctly specified. Estimates were computed over 1000 replicate datasets.}
    \label{fig:param_estimates}
\end{figure}

Table~\ref{tab:1to3} displays root-$n$ adjusted biases and standard errors of the blip parameter estimators across the four model specifications and sample sizes. Figure \ref{fig:param_estimates}  shows the empirical distributions of the blip parameter estimates across replications for $n_{\text{train}} = 1000$. The figure for the larger sample size can be found in Appendix A. As expected, we have consistent estimation of the blip parameters when either the treatment-free or weighting models are correctly specified. However, misspecifying the treatment-free component incurs a larger penalty in terms of efficiency than misspecifying the weighting models, as is evidenced by the larger variances and small sample bias of the estimators under model (iii), as opposed to those under model (ii).

Table~\ref{tab:1to3} shows the estimated POTs and values for the estimated weighted and greedy ITRs as well as the values for the uniform and oracle regimes across models for both sample sizes. For these data generating mechanisms, models (ii)-(iv) all behave very similarly in terms of POT and value, with values that are very close to that of the oracle regime. As expected, model (i), for which the blip estimators are inconsistent, performs poorly.

\begin{figure}[H]
    \centering
    \includegraphics[scale = 0.53]{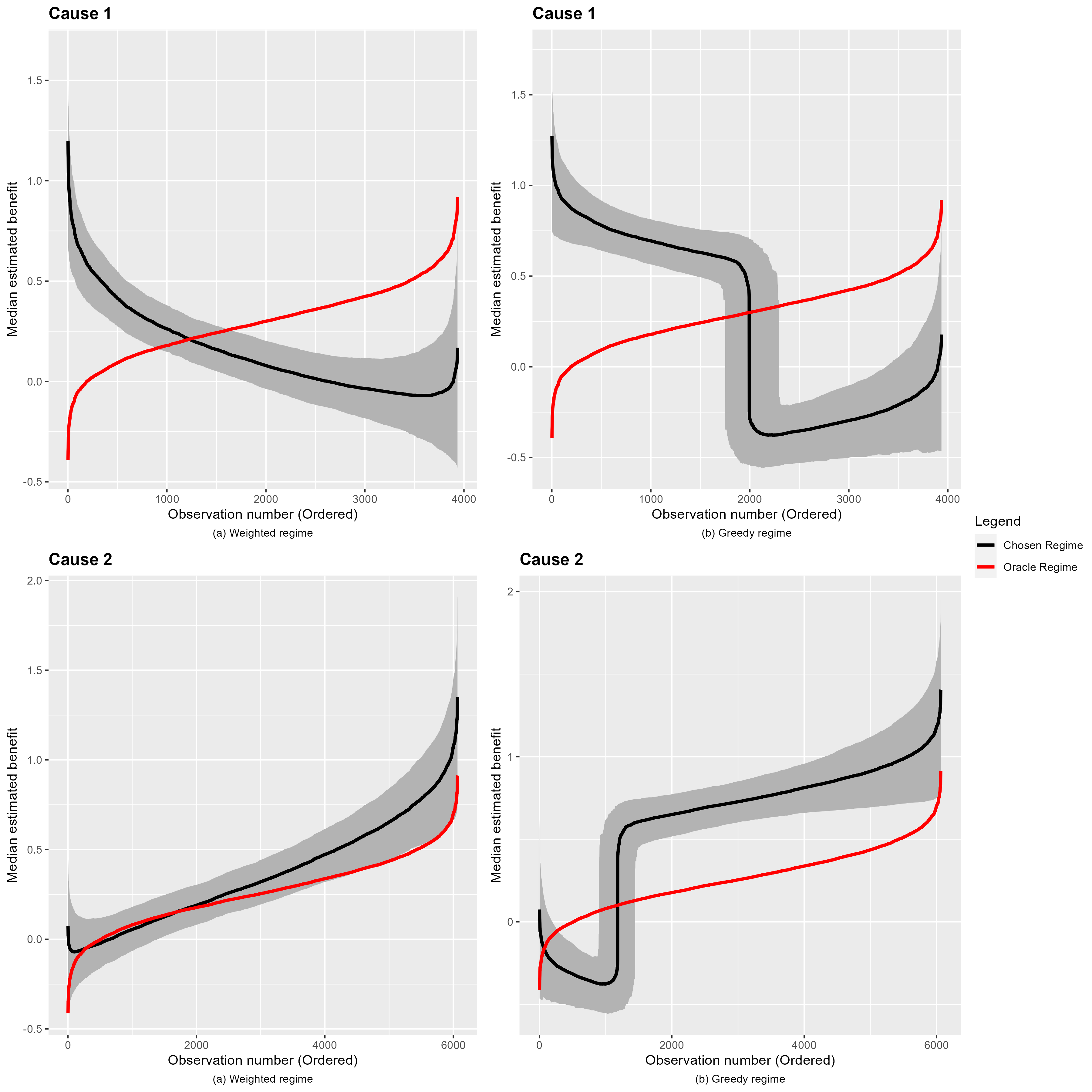}
    \caption{Median estimated benefit for the weighted (left column) and greedy (right column) regimes along with the benefit of the oracle regime for setting 1(a) and model (i), for causes $K=1$ (top row) and $K=2$ (bottom row), $n_{\text{train}} = 1000$. Benefit curves were evaluated on the test set, with subjects ordered by their increasing benefit. Estimates were computed over 1000 replicate datasets.}
   \label{fig:b1_model1}
 \end{figure}

In Figure \ref{fig:b1_model1}, we present the median estimated benefit of the greedy ITR and the weighted ITR for subjects in the test set, along with pointwise $95\%$ CIs, for model (i) in setting $1$(a). The plot is stratified by true cause of failure and subjects are ordered by their benefit in increasing order. Within the same figure, we plot the benefit for the oracle regime, obtained from the true value of the blip parameters. Interpreting the benefit plots is relatively straightforward: a given ITR $d$ prescribes treatment whenever $\mathcal{B}_d(\bm{x}) > 0$, i.e., when the plotted line is above the horizontal line at $0$. An estimated ITR prescribes the wrong treatment whenever its benefit curve has the opposite sign of the benefit of the oracle regime, i.e., when the black and red lines are on opposite sides of the horizontal line at $0$. Generally, we expect a given ITR to perform better in terms of POT and value, if its estimated benefit function is closer to the benefit of the oracle regime. This can be easily deduced from the proximity of the lines in the benefit plot. As previously noted, model (i) performs worse than model (iv) in terms of value and POT, with the benefit curves of the estimated regimes being far from the benefit of the oracle regime whenever all models are misspecified. Benefit plots for all other simulation settings can be found in Appendix A.

\subsection{Comparison of optimality criteria and resulting rules}
\label{subsec:comparison}

The second set of simulations was conducted in order to compare the behaviors of the greedy rule and the weighted rule under different data-generating mechanisms.
We considered the settings defined by the following values of the blip parameters, with $n_{\text{train}} = 1000$ and $r = 50$:
\begin{align*}
    &\text{Setting } 2: (\psi_{11}, \psi_{12}) = (3, -0.5), \quad (\psi_{21}, \psi_{22}) = (-1, 0.2).\\
    &\text{Setting } 3: (\psi_{11}, \psi_{12}) = (-0.5, -0.7), \quad (\psi_{21}, \psi_{22}) = (0.1, 0.08).
\end{align*}

Table \ref{tab:1to3} showcases the biases and SEs of the blip parameter estimators for the two settings as well as the values and POTs of the estimated regimes. In setting $2$, the weighted regime yields notably a higher value than the greedy regime, while having a moderately smaller value of POT. Since the blip parameters for both causes have relatively similar magnitudes and opposite signs, the weighted ITR's averaging of the two blips allows the benefit to be close to zero when the failure type probability is near $0.5$. This allows the weighted regime to switch from one cause to the other much more quickly than the greedy regime, yielding the superior performance that we observe. 

In the setting $3$, the weighted regime performs slightly better than the greedy regime in terms of value, at the cost of a much lower POT. As a result, the greedy ITR might be preferable in this setting on the basis of overall performance. In this case, the magnitudes of the blip parameters for the second cause are smaller than those for the first cause, and so averaging the two blips yields a benefit that is much closer to the benefit associated with the first cause. In other words, the weighted regime is ``stuck" on the first cause, a problem that the greedy regime does not share. 

In the additional simulation settings, we observe that the performance of both the greedy and the weighted ITRs is robust to variations in the data generating mechanism, with performance that matches the results presented above.

\subsection{Comparison with standard methods}

The final set of simulations compares our proposed method to standard approaches used in the competing risks literature. 

\subsubsection{Cause-specific ITR}
The first approach, used in \textit{cause-specific} analyses, censors the competing event and estimates the optimal ITR based on one cause of interest \citep{Choi2019}. Without loss of generality, we assume that the targeted cause of failure is cause 1. The cause-specific ITR is then estimated via DWSurv with outcome variable the time to failure from cause 1, and failures due to cause 2 treated as censored.
We generate data according to the data generating mechanism of Subsection \ref{subsec:dgm}, with the following blip parameters and cause model:
\begin{align*}
    \text{Setting } 10:\quad  &(\psi_{11}, \psi_{12}) = (1, -0.5), \quad (\psi_{21}, \psi_{22}) = (-3, 0.2),\\
    &\quad K_{ij} \sim \text{Bernoulli}\Bigl(\text{expit}(2.5 + X_{i1})\Bigr) + 1,
\end{align*}
which corresponds to a study population with approximately 10\% of patients failing from cause 1.
In this setting, the two causes of failure have vastly different associated ITRs. Allocating treatment based on only one cause of failure can then yield disastrous performance when the treatment is harmful for those failing from the second cause, especially when the second cause is much more prevalent. 

Blip parameter estimates and measures of uncertainty, as well as the  metrics for our proposed ITRs and the cause-specific ITR are presented in Appendix A. The cause-specific ITR yields POT and value estimates of $0.19$ and $-0.65$, respectively. As expected, the cause-specific ITR is nearly optimal for those failing from the targeted cause, but is far from optimal for those who experience the alternative cause of failure. Since the second cause is more prevalent, the approach yields overall performance worse than even the uniform regime, which has an estimated value of $0.61$. The weighted and greedy ITRs perform very well in this setting, both with POT and value estimates of $0.90$ and $1.87$, respectively.

\subsubsection{Composite outcome ITR}
\label{subsubsec:composite}
The second approach constructs an ITR based on a composite endpoint which includes all event types. This amounts to ignoring distinct causes of failure and estimating one set of blip parameters based on all observed data. Heuristically, this corresponds to a form of weighting, with the combined set of blip parameters being a weighted average of the blip parameters associated to each cause, weighted by the prevalence of each cause. The composite ITR is estimated via DWSurv, with  the outcome being the time to any event. 

To illustrate this, we consider Setting $4$, where the greedy regime outperforms the weighted regime. In this setting, the composite ITR yields POT and value estimates of $0.49$ and $1.54$ respectively, nearly identical to those of the weighted regime. The composite ITR shares the same flaws as the weighted regime and is similarly surpassed by the greedy regime in this setting.  In such situations, a composite ITR provides sensible results; however, the added flexibility of being able to choose between the weighted and greedy regimes based on the objective of the analysis is an advantage of our proposed approach.

\section{Application: OPTN kidney transplant data}
\label{sec:application}

\subsection{Background: Donor HCV status}
In this section, we illustrate our proposed method via application to the kidney transplant data from the OPTN database. We analyze data from patients enrolled on the kidney transplantation waiting list from January 1, 2001 to December 31, 2022. The main exposure of interest is the donor's HCV status, with 0 denoting an HCV-negative donor and 1 denoting a donor living with HCV. We seek to maximize time to either graft failure (cause 1), or death with a functioning graft (cause 2). We include donor age, donor type (living or dead), and recipient HCV status as possible tailoring variables, with donor age scaled by 10 years. Other key variables, including those used for the treatment-free component of the outcome model, are presented in Appendix B. For the sake of simplicity, we conducted a complete-case analysis, removing patients with missing event indicators, missing survival times or missing covariates. We emphasize that the goal of this analysis is to illustrate our methodology rather than to provide accurate treatment recommendations. 

\subsection{Data and analytic models}
Observations in our dataset were clustered by transplant centers, with a total of 251 distinct centers. We used an exchangeable working correlation structure. Due to large cluster sizes, we used a single iteration of the algorithm in Subsection \ref{subsec:estimation} with initial correlation estimates obtained by fitting a weighted linear regression under independence. For the exchangeable structure, the inverse of the working correlation matrix can be derived analytically, leading to greater computational efficiency \citep{Lipsitz2017}. Weights as in (\ref{eq:overlap}) were used for fitting the weighted GEE.  Confidence intervals for blip parameters were obtained via nonparametric cluster bootstrap with $B = 1000$ replications. Logistic regression was used for the treatment, censoring and cause models; details of model specification are located in Appendix B. All outcome models were fitted in R using the \texttt{geem} package. 

Patient characteristics are presented in Appendix B. Our final dataset was made up of 311,474 patients, with 15.9\% suffering from graft failure, 16.1\% dying with a functioning graft and 68\% with censored outcomes. Transplants with HCV positive donors were rare within the overall study population, constituting only 3.6\% of all transplants. This proportion was larger within HCV positive recipients, at 31\%.  

\subsection{Results}
Table \ref{tab:data_analysis_estimates} shows blip parameter estimates for each cause of failure along with associated measures of uncertainty.  For both causes, the one-step procedure yielded an estimated within-cluster correlation of around 5\%.  For patients having suffered from graft failure, receiving a kidney from an HCV positive donor over an HCV negative donor has a negative effect on survival time; however, the recipient being HCV positive and the transplant coming from a living donor both significantly attenuate the negative effect. Effect estimates were similar for patients who died with a functioning graft, with the only difference being the statistically significant negative effect of donor HCV status on survival time. Donor age had no significant tailoring effect, either statistically or clinically. 

The two regimes provided nearly identical performance, with an estimated POT of around $0.94$ and an estimated value of $6.96$. Since the blip parameters are fairly similar for both causes, either regime appears to be a good choice. Benefit plots for the weighted and greedy ITRs can be found in Appendix B. For most patients, the estimated ITRs recommend transplants with HCV-negative donors over transplants with HCV-positive donors, which is consistent with previous literature. However, for both regimes, a relatively large portion of patients have estimated benefits for which the $95$\% bootstrap CI include $0$: $20$\% and $28$\% of subjects respectively. For these patients, there is no apparent harm to using an HCV-positive donor over an HCV-negative donor as both yield similar outcomes. We note that the observed treatment allocation also results in an estimated value of $6.96$; however, only $3.63\%$ of patients received kidneys from an HCV-positive donor. Under either of our proposed regimes, a much larger proportion of the study population could have been allocated kidneys from an HCV-positive donor without impacting their overall survival. 

Additionally, we conducted the same analysis using the cause-specific and composite outcome approaches. Parameter estimates and benefit plots for these two analyses can be found in Appendix B.  The composite outcome analysis yielded estimates consistent with those obtained in the main analysis and the resulting ITR exhibited performance similar to that of the weighted regime, as was observed in Subsection \ref{subsubsec:composite} of the simulations. The cause-specific analysis also yielded reasonable point estimates; however, the associated standard errors were much larger than the equivalent standard errors in the main analysis. In this case, the artificial censoring triggered by the cause-specific analysis leads to greater variability in the estimated weights. An unweighted analysis could be performed to remedy this issue, at the cost of increased sensitivity to model misspecification. Furthermore, the cause-specific ITR performs worse than all the previously mentioned regimes, with an estimated POT and value of $0.79$ and $6.79$, respectively. Although the estimated blip parameters are similar across causes, they are not identical; in this case, only allocating treatment based on one failure type results in sub-optimal performance.

\section{Conclusion} \label{sec:conclusion}

In this paper, we proposed a framework for optimal ITR estimation in the competing risk setting which accounts for both treatment effect heterogeneity and the uncertainty related to the unknown failure type. Estimates of treatment effects are obtained by solving doubly-robust GEEs, with weights adjusted for non-randomized treatment assignment and censoring. Treatment effect estimates can then be combined in various ways to yield optimal ITRs which target different criteria. For example, we introduced the weighted and greedy ITRs, which target overall survival integrated over all failure types and survival tied to the most likely failure type, respectively. The choice of an optimal ITR should mainly be driven by subject matter expertise and the nature of the research question, but metrics such as the value and the POT may be used to compare the performance of the chosen treatment rule to that of alternative regimes. 

The main advantage of our approach is that it explicitly incorporates the uncertainty regarding the failure type into the definition of the optimal ITR, rather than implicitly assuming that the cause of failure is known at treatment allocation. Although one can repeat a naive cause-specific analysis for every event type separately, there is limited formal methodology for combining these results into a single optimal treatment rule. Alternatively, as discussed in Subsection \ref{subsubsec:composite}, the use of a composite outcome can yield similar performance to that of the weighted ITR, at least in the case $K=2$; consequently, it also performs poorly where the weighted regime does. In these cases, one could instead use the greedy rule, which was shown to yield better performance in certain scenarios where the weighted rule is less effective. Another strength of our approach is the AFT specification, which allows for more intuitive modelling, as treatment effects can be directly interpreted in terms of the survival time, which is not the case for hazard-based models.

One weakness of our proposed approach is that it requires the relatively strong assumption that the cause of failure $K$ is conditionally independent of the joint distribution of the treatment and censoring mechanism given observed covariates, i.e., Assumptions 3 and 5. Conditional independence of the failure type and treatment assignment may not hold if, for example, the treatment induces side-effects that can cause patients to fail before they experience the event type of interest. To remedy this particular problem, we can derive alternative definitions of the optimal ITR under the weaker assumption of sequential ignorability, leading to the optimal rule $ d_{\text{w}}(x) =  \arg \max_a \sum_{k=1}^\kappa \varphi_k(\bm{x}, a) Q_k(\bm{x}, a), \; \bm{x} \in \mathcal{X}$, where $\varphi_k(\bm{x}, a) = P(K = k| \bm{X} = \bm{x}, A = a)$ and $Q_k(\bm{x}, a) = \mathbb{E}[f(T)| \bm{X} = \bm{x}, A = a, K = k]$. See Appendix C for details. The optimal ITR can then be directly estimated from the data by positing regression models for the transformed survival times and the failure type. However, this approach is singly-robust and thus is intrinsically more sensitive to model misspecification. Strategies to improve the robustness of this approach will be explored in future work. Furthermore, if we do not assume that the censoring mechanism is conditionally independent of the failure type, the algorithm outlined in Subsection \ref{subsec:estimation} would require the specification of the censoring probability for each subject conditional on their failure type. Since the failure type is unobserved for censored subjects, this would require leveraging the observed data to impute missing failure types. As noted in Subsection \ref{subsec:mixture}, several different methods could be used in order to achieve this goal. 

We leveraged our methodological developments to assess the important question of kidney allocation with HCV-positive donors using data from the OPTN. We found statistically significant tailoring effects for the recipient HCV status and the donor type. The resulting ITRs were then used to identify patients for whom kidney transplants with HCV-positive donors did not yield inferior outcomes to transplants with HCV-negative donors. Our analysis suggests that a large portion of patients in the OPTN data who received HCV-negative kidneys would see their overall survival unchanged if they were instead given a kidney from an HCV-positive donor. The estimated treatment rules could then be used to allocate kidneys for future transplant candidates, allowing for a more efficient use of kidneys from HCV-positive donors. However, our analysis was subject to certain limitations. First, we performed a complete case analysis, which could lead to bias if the variables concerned were not missing completely at random. Furthermore, our choice of tailoring variables was mainly based on substantive clinical knowledge and limited pre-screening of effect sizes. A more principled data-driven variable selection procedure, such as the one utilized in penalized dWOLS \citep{bian2021}, could be used to narrow down the set of candidate predictors. 

In spite of these limitations, our analysis had several strengths including a large sample size, appropriate accounting of confounding, censoring, and clustering in addition to robustness against model misspecification. The work shown here represents the first ITR approach to formalizing decision-making for kidney-allocation with a goal to minimizing harm in the context of potentially sub-optimal donor organs, and has revealed that a much larger group of recipients could be offered organs from donors living with HCV without evidence that this would result in recipient harm.


\begin{table}[H]
    \centering
     \caption{Root-$n$ adjusted bias and standard errors for blip parameter estimators, as well as ITR metrics for settings $1$-$3$. Estimates and metrics for setting $1$(a) and $1$(b) were computed across $4$ model specifications: (i) outcome and weighting models misspecified, (ii) outcome model correctly specified, (iii) weighting models correctly specified, (iv) all models correctly specified.  Estimates were computed using 1000 replicate datasets. Regimes are identified with $d_w$, $d_g$, $d_o$, and $d_u$ representing the weighted, greedy, oracle and uniform regimes, respectively.\\ }
    \begin{tabular}{|l|c|c|c|c|c|c|c|c|c|c|c|}
    \hline
        \multirow{3}{*}{} & \multirow{3}{*}{$\bm{\psi}$} &\multicolumn{10}{|c|}{Setting}\\
        & & \multicolumn{4}{|c|}{1(a)}&  \multicolumn{4}{|c|}{1(b)} &2 &3 \\
        & & i& ii& iii& iv& i& ii& iii&iv & - & -\\
        \hline
         $\sqrt{n}\; \times$ Bias & $\psi_{11}$& -22.41& -0.04 &-0.27 &-0.07 &-50.09 &0.09 &0.09 &0.13 & 0.10& -0.07\\
         & $\psi_{12}$ &0.30 &-0.08 &-0.12 &-0.10 &1.00 &0.10 &0.12 &0.12 &-0.00 &-0.10 \\
         & $\psi_{21}$ &14.96 &0.06 &0.11 &0.05 &33.58 &0.01 &0.05 &-0.01 &0.07 &0.05 \\
         & $\psi_{22}$ &0.06 &0.08 &0.12 &0.08 &0.08 &0.13 &0.25 &0.19 &0.14 &0.08 \\
         \hline
         $\sqrt{n}\; \times$ SE & $\psi_{11}$&3.23 &2.79 &3.78 &3.29 &3.17 &2.88 &4.14 &3.53 &3.49 &3.29 \\
         & $\psi_{12}$ &3.30 &2.52 &3.98 &3.37 &3.38 &2.59 &4.18 &3.50&3.49 &3.37 \\
         & $\psi_{21}$ &2.00 &2.21 &2.67 &2.54 &1.99 &2.18 &2.72 &2.58 &2.60 &2.54 \\
         & $\psi_{22}$ &2.72 &2.43 &3.32 &3.08 &2.61 &2.34 &3.29 &2.98 &3.12 &3.08 \\
         \hline
         \multicolumn{2}{|c|}{POT($\hat{d}_{\text{w}}$)}& 0.84& 0.93 &0.93 &0.93 &0.83 &0.93 &0.93 &0.93 & 0.56& 0.51 \\
         \multicolumn{2}{|c|}{POT($\hat{d}_{\text{g}}$)}&0.73 &0.92 &0.92 &0.92 &0.72 &0.93 &0.93 &0.93 &0.69 &0.70 \\
         \hline
         \multicolumn{2}{|c|}{V($\hat{d}_{\text{w}}$)}&1.73 &1.81 &1.81 &1.81 &1.70 &1.76 &1.76 &1.76 &2.30 &1.61 \\
         \multicolumn{2}{|c|}{V($\hat{d}_{\text{g}}$)}&1.72 &1.81 &1.79 &1.81 &1.68 &1.76 &1.76 &1.76 &2.17 &1.55 \\
         \hline
         \multicolumn{2}{|c|}{V($d_{\text{o}}$)}&1.81 &1.81 &1.81 &1.81 &1.77 &1.77 &1.77 &1.77 &2.81 &1.68 \\
         \multicolumn{2}{|c|}{V($d_{\text{u}}$)}&1.67 &1.67 &1.67 &1.67 &1.62 &1.62 &1.62 &1.77 &1.89 &1.54 \\
         \hline
    \end{tabular}
   
    \label{tab:1to3}
\end{table}

\begin{table}[H]
    \centering
    \caption{Blip parameter estimates and associated 95\% bootstrap confidence intervals (CI) for both causes of failure in the OPTN data on $n = 311,474$ individuals from $251$ centers, based on kidney transplantations carried out from January 1, 2001 to December 31, 2022. Nonparametric cluster bootstrap CIs were computed using $B=1000$ bootstrap replicates. Donor and recipient variables are prefixed by Don and Rec, respectively.\\}
    \begin{tabular}{|c|c|c|c|c|}
    \hline
     \multirow{2}{*}{Parameters}& \multicolumn{2}{|c|}{Graft Failure} & \multicolumn{2}{|c|}{Death with functioning graft}\\
     &  Estimate & 95\% CI & Estimate & 95\% CI\\
    \hline
    DonHCV & $-0.86$& $(-2.04, 0.04)$ & $-1.44$& $(-2.30, -0.51)$ \\
    DonHCVxRecHCV & $0.90$& $(0.55, 1.32)$& $1.21$& $(0.97, 1.47)$\\
    DonHCVxDonType & $1.14$& $(0.61, 1.71)$ & $0.40$& $(-0.21, 1.01)$\\
    DonHCVxDonAge & $1.55\text{x} 10^{-5}$& $(-1.36\text{x}10^{-3}, 1.82\text{x} 10^{-3})$& $4.32\text{x}10^{-4}$& $(-1.00\text{x} 10^{-3}, 1.73\text{x}10^{-3})$\\
    \hline
    \end{tabular}
    \label{tab:data_analysis_estimates}
\end{table}

\bigskip
\begin{center}
{\large\bf SUPPLEMENTARY MATERIAL}
\end{center}

\begin{description}

\item[Appendix A, B, and C:] Appendix A contains additional simulations and figures. Appendix B provides supplementary information regarding the data analysis. Appendix C presents proofs for the main results of the paper. (.pdf files)

\end{description}

\bibliography{refs}

\end{document}


\pagenumbering{gobble}

\section*{Appendix A}

\subsection*{Plots for simulations}

\begin{figure}[H]
     \centering
     \includegraphics[scale = 0.30]{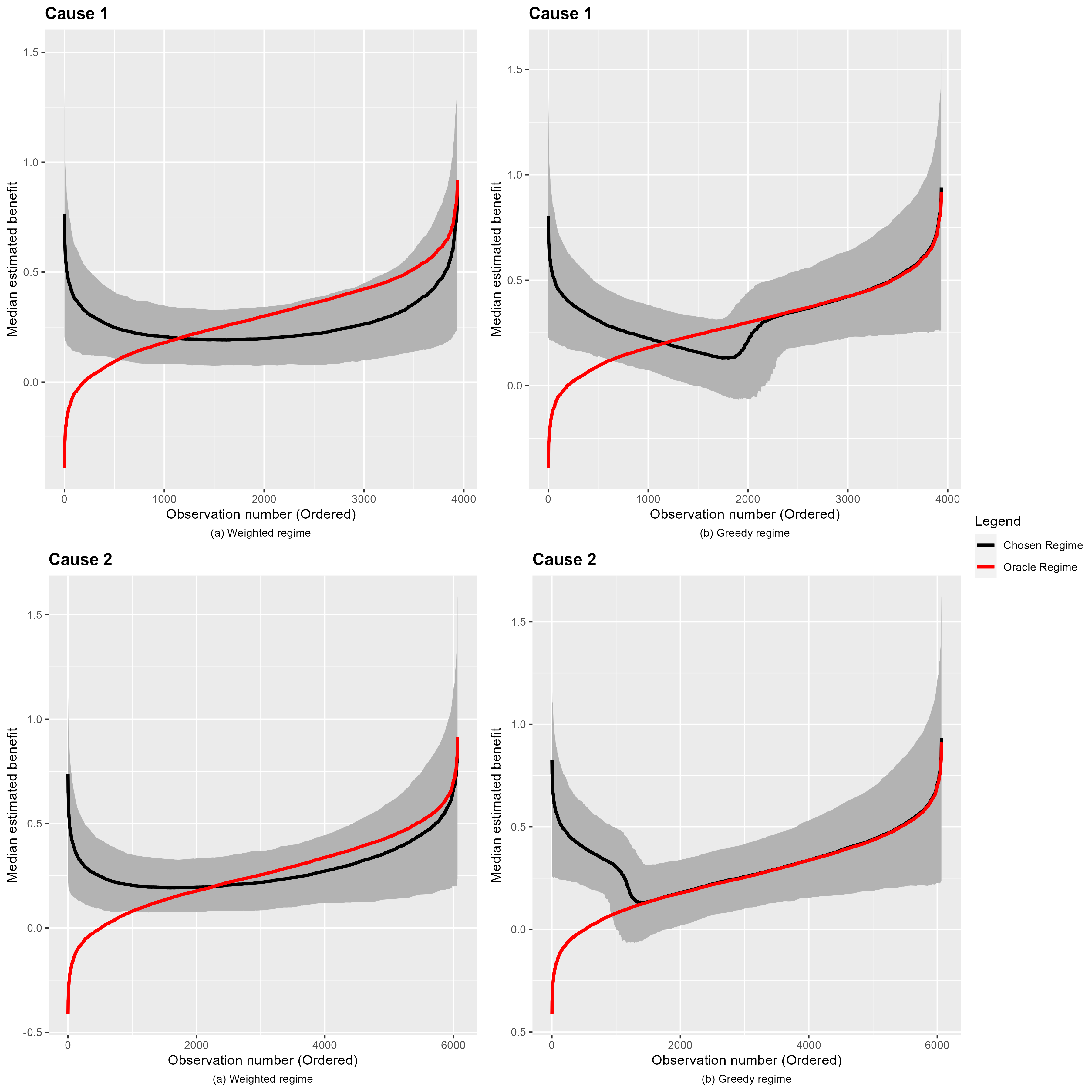}
     \caption*{Figure A1: Median estimated benefit for the weighted (left column) and greedy (right column) regimes along with the benefit of the oracle regime for setting 1(a) and model (iv), for causes $K=1$ (top row) and $K=2$ (bottom row), $n_{\text{train}} = 1000$. Benefit curves were evaluated on the test set, with subjects ordered by their benefit in increasing order. Estimates were computed over 1000 replicate datasets.}
     \label{fig:b1_model2}
\end{figure}

 \begin{figure}[H]
    \centering
    \includegraphics[scale = 0.30]{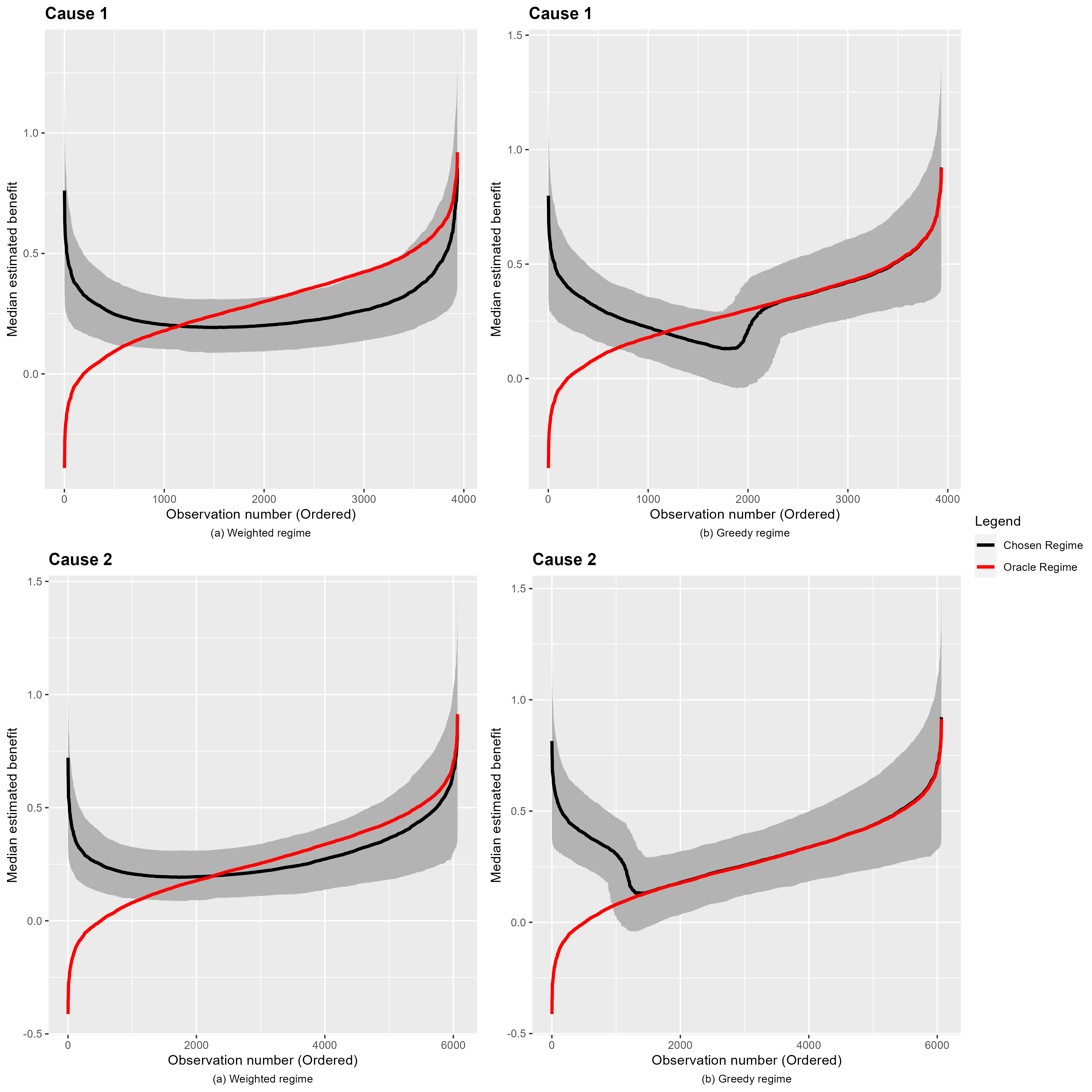}
    \caption*{Figure A2:
    Median estimated benefit for the weighted (left column) and greedy (right column) regimes along with the benefit of the oracle regime for model setting 1(a) and model (ii), for causes $K=1$ (top row) and $K=2$ (bottom row), $n_{\text{train}} = 1000$. Benefit curves were evaluated on the test set, with subjects ordered by their benefit in increasing order. Estimates were computed over 1000 replicate datasets.}
 \end{figure}

  \begin{figure}[H]
    \centering
    \includegraphics[scale = 0.30]{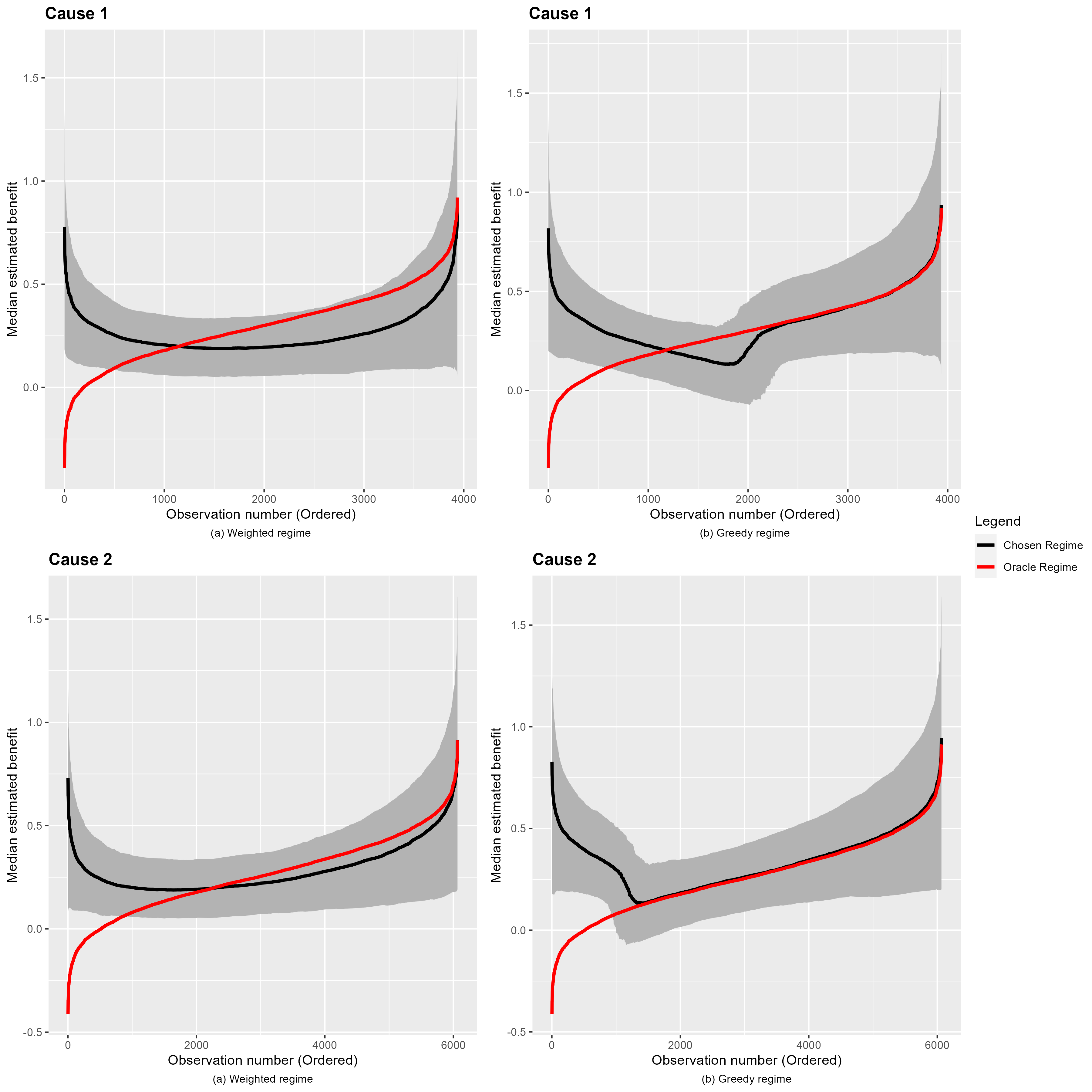}
    \caption*{Figure A3: Median estimated benefit for the weighted (left column) and greedy (right column) regimes along with the benefit of the oracle regime for setting 1(a) and model (iii), for causes $K=1$ (top row) and $K=2$ (bottom row), $n_{\text{train}} = 1000$. Benefit curves were evaluated on the test set, with subjects ordered by their benefit in increasing order. Estimates were computed over 1000 replicate datasets. }
 \end{figure}

\begin{figure}[H]
    \centering
    \includegraphics[scale = 0.30]{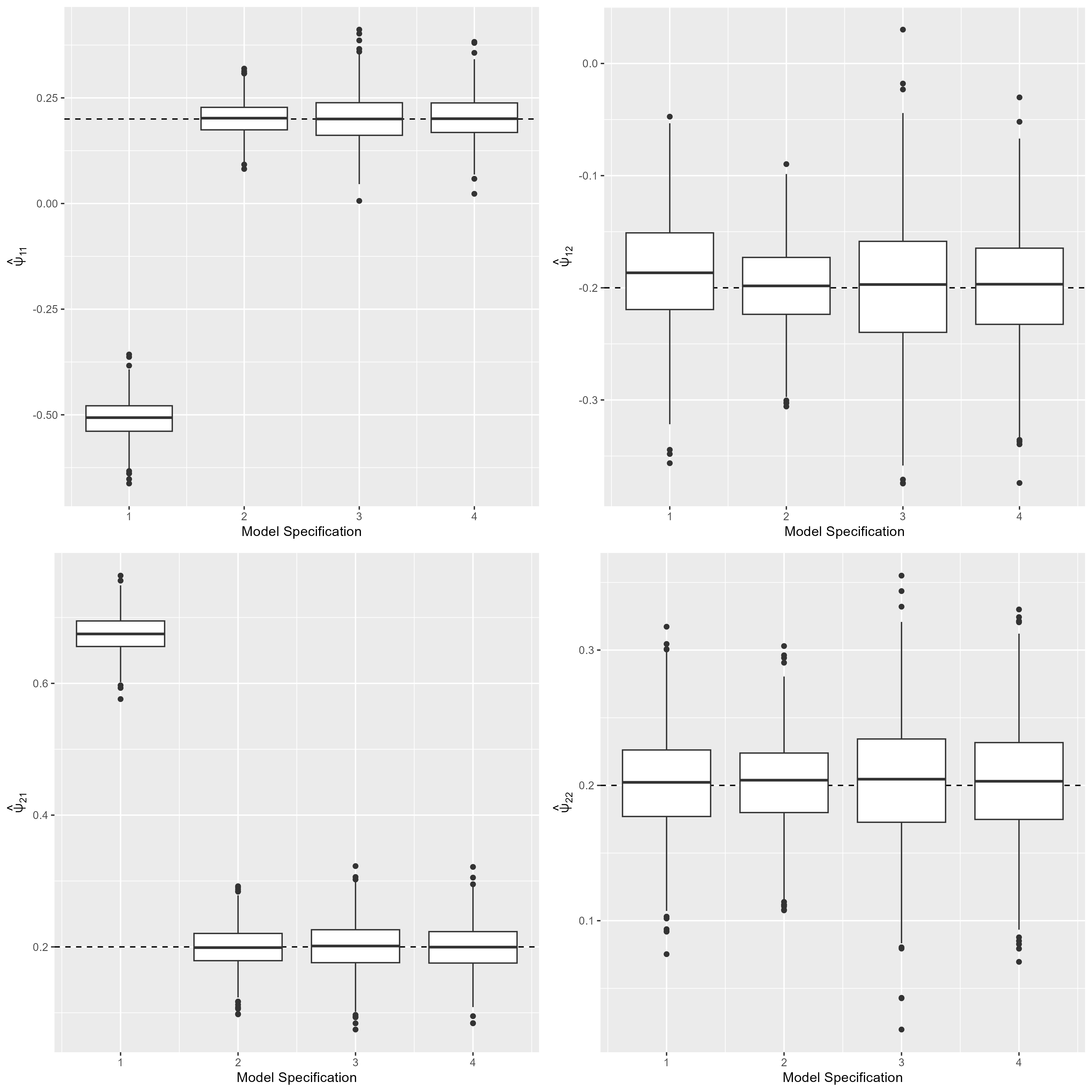}
    \caption*{Figure A4: Empirical distribution of blip parameter estimates obtained via the weighted AFT-GEE model with sample size $n_{\text{train}} = 5000$ in setting 1(b) across 4 scenarios: (i) outcome and weighting models misspecified, (ii) outcome model correctly specified, (iii) weighting models correctly specified, (iv) all models correctly specified. Estimates were computed over 1000 replicate datasets.}
\end{figure}

  \begin{figure}[H]
    \centering
    \includegraphics[scale = 0.30]{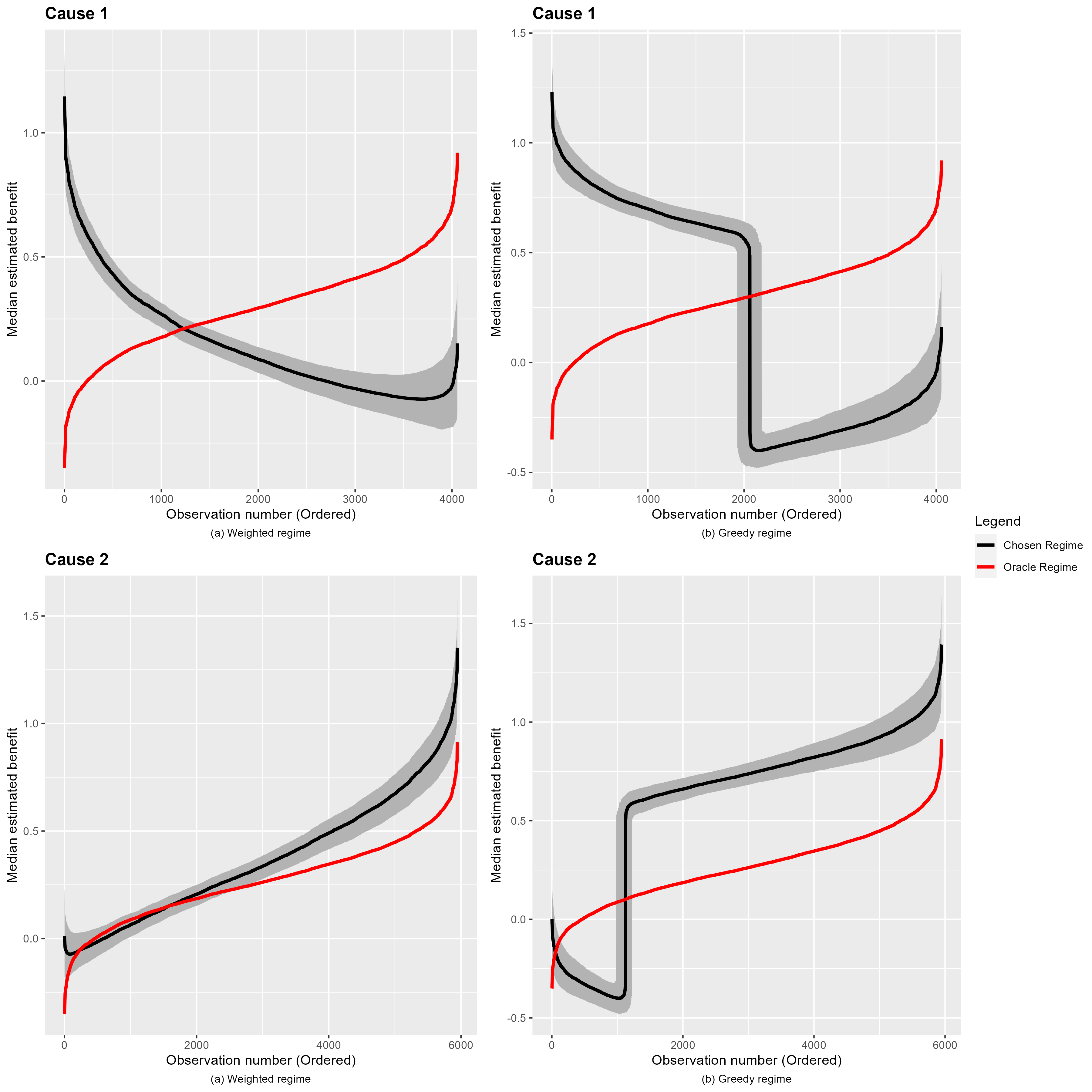}
    \caption*{Figure A5: Median estimated benefit for the weighted (left column) and greedy (right column) regimes along with the benefit of the oracle regime for setting 1(b) and model (i), for causes $K=1$ (top row) and $K=2$ (bottom row), $n_{\text{train}} = 5000$. Benefit curves were evaluated on the test set, with subjects ordered by their benefit in increasing order. Estimates were computed over 1000 replicate datasets. }
 \end{figure}

\begin{figure}[H]
    \centering
    \includegraphics[scale = 0.30]{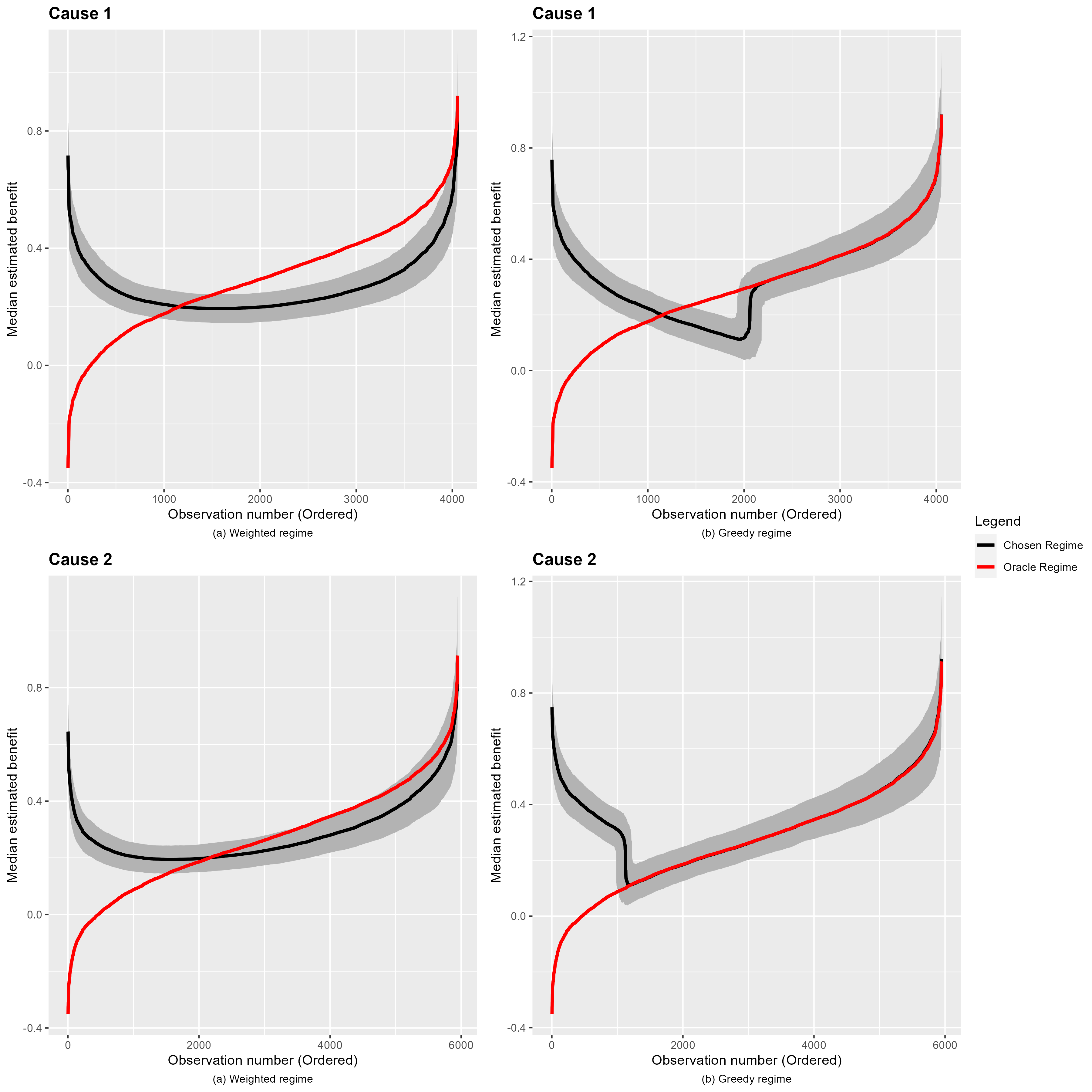}
    \caption*{Figure A6: Median estimated benefit for the weighted (left column) and greedy (right column) regimes along with the benefit of the oracle regime for setting 1(b) and model (ii), for causes $K=1$ (top row) and $K=2$ (bottom row), $n_{\text{train}} = 5000$. Benefit curves were evaluated on the test set, with subjects ordered by their benefit in increasing order. Estimates were computed over 1000 replicate datasets. }
 \end{figure}
 
 \begin{figure}[H]
    \centering
    \includegraphics[scale = 0.30]{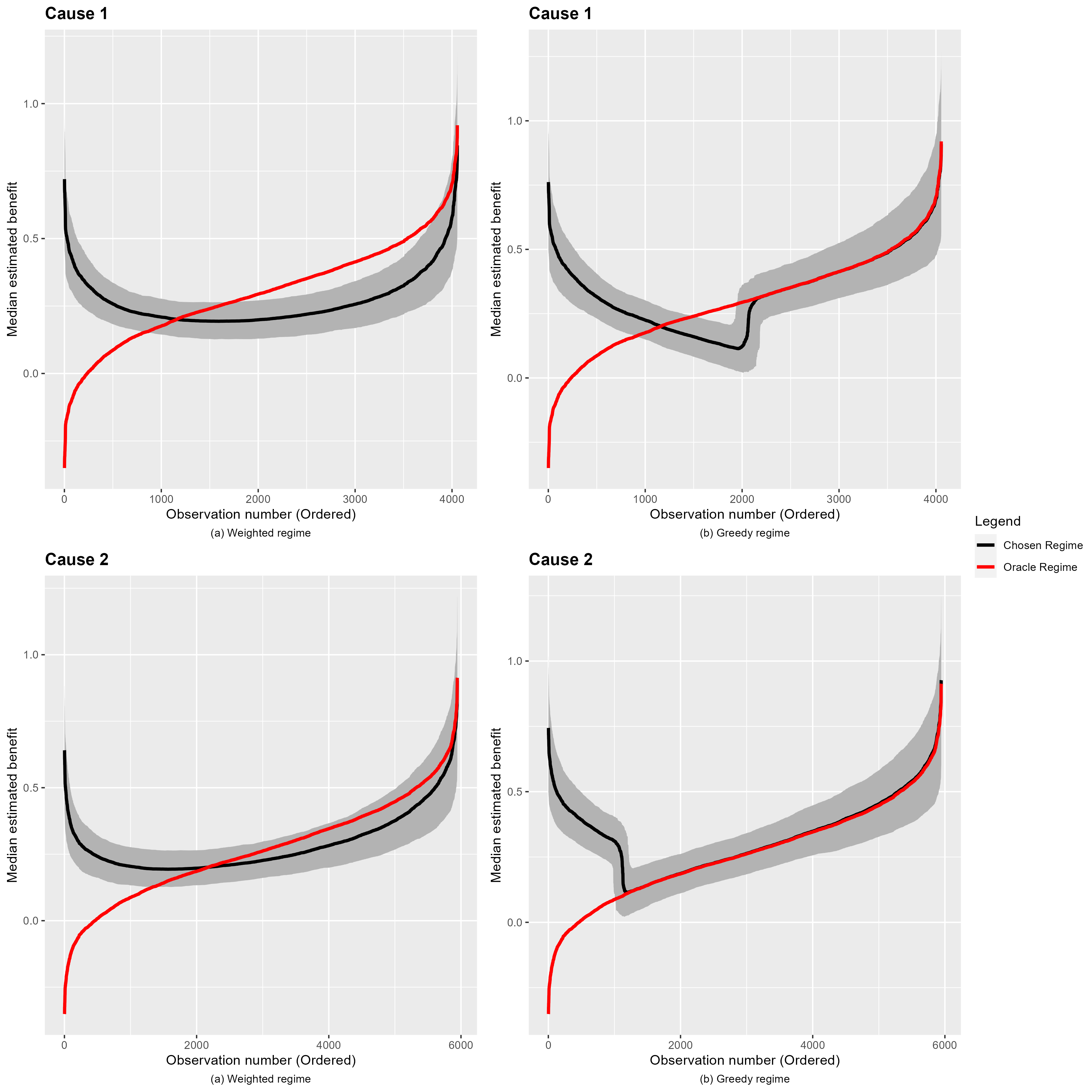}
    \caption*{Figure A7: Median estimated benefit for the weighted (left column) and greedy (right column) regimes along with the benefit of the oracle regime for setting 1(b) and model (iii), for causes $K=1$ (top row) and $K=2$ (bottom row), $n_{\text{train}} = 5000$. Benefit curves were evaluated on the test set, with subjects ordered by their benefit in increasing order. Estimates were computed over 1000 replicate datasets.}
 \end{figure}

 \begin{figure}[H]
    \centering
    \includegraphics[scale = 0.30]{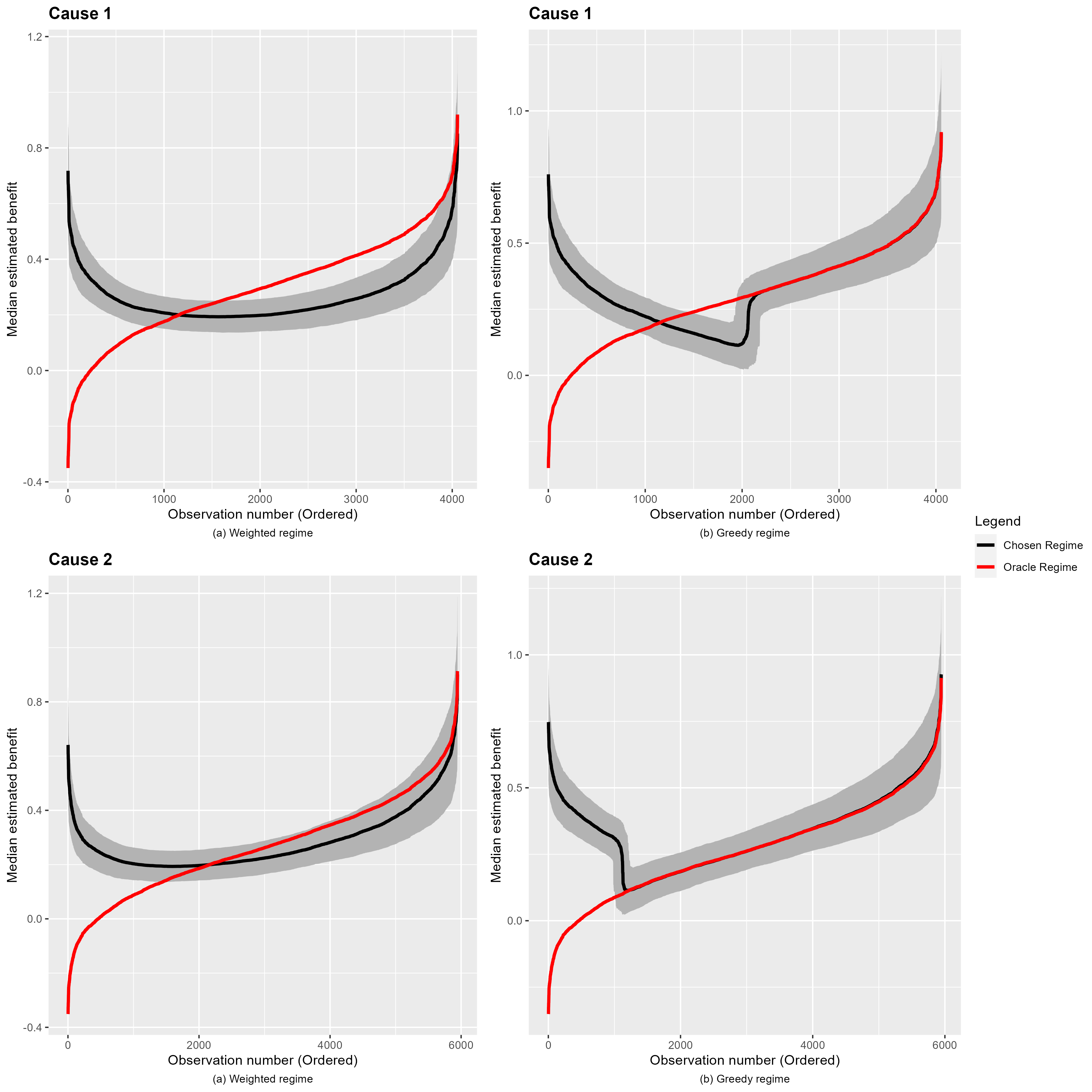}
    \caption*{Figure A8: Median estimated benefit for the weighted (left column) and greedy (right column) regimes along with the benefit of the oracle regime for setting 1(b) and model (iv), for causes $K=1$ (top row) and $K=2$ (bottom row), $n_{\text{train}} = 5000$. Benefit curves were evaluated on the test set, with subjects ordered by their benefit in increasing order. Estimates were computed over 1000 replicate datasets.}
 \end{figure}

\begin{figure}[H]
    \centering
    \includegraphics[scale = 0.30]{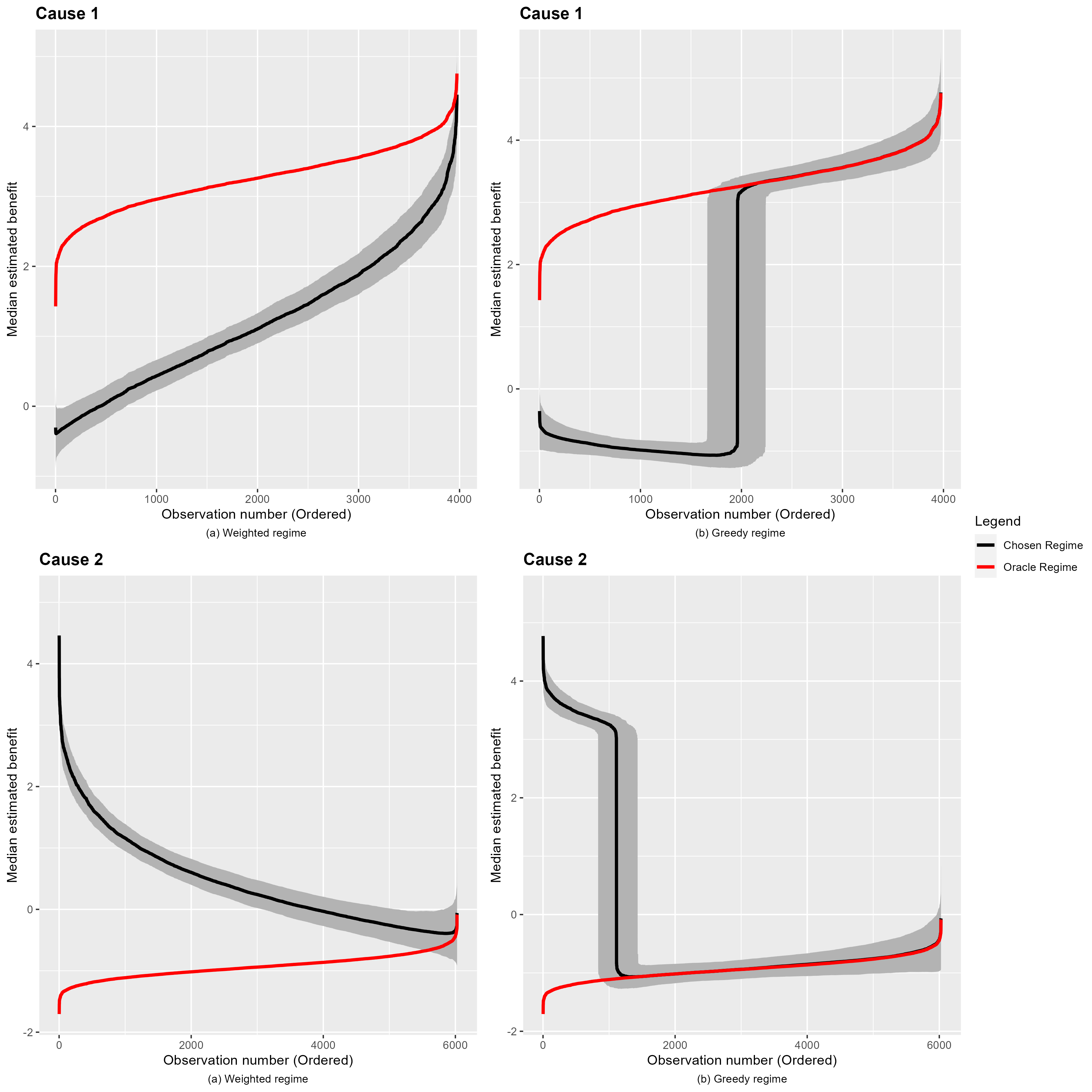}
    \caption*{Figure A9: Median estimated benefit for the weighted (left column) and greedy (right column) regimes along with the benefit of the oracle regime for setting 2, for causes $K=1$ (top row) and $K=2$ (bottom row). Benefit curves were evaluated on the test set, with subjects ordered by their benefit in increasing order. Estimates were computed over 1000 replicate datasets.}
   \label{fig:comparison_s1}
 \end{figure}

 \begin{figure}[H]
    \centering
    \includegraphics[scale = 0.30]{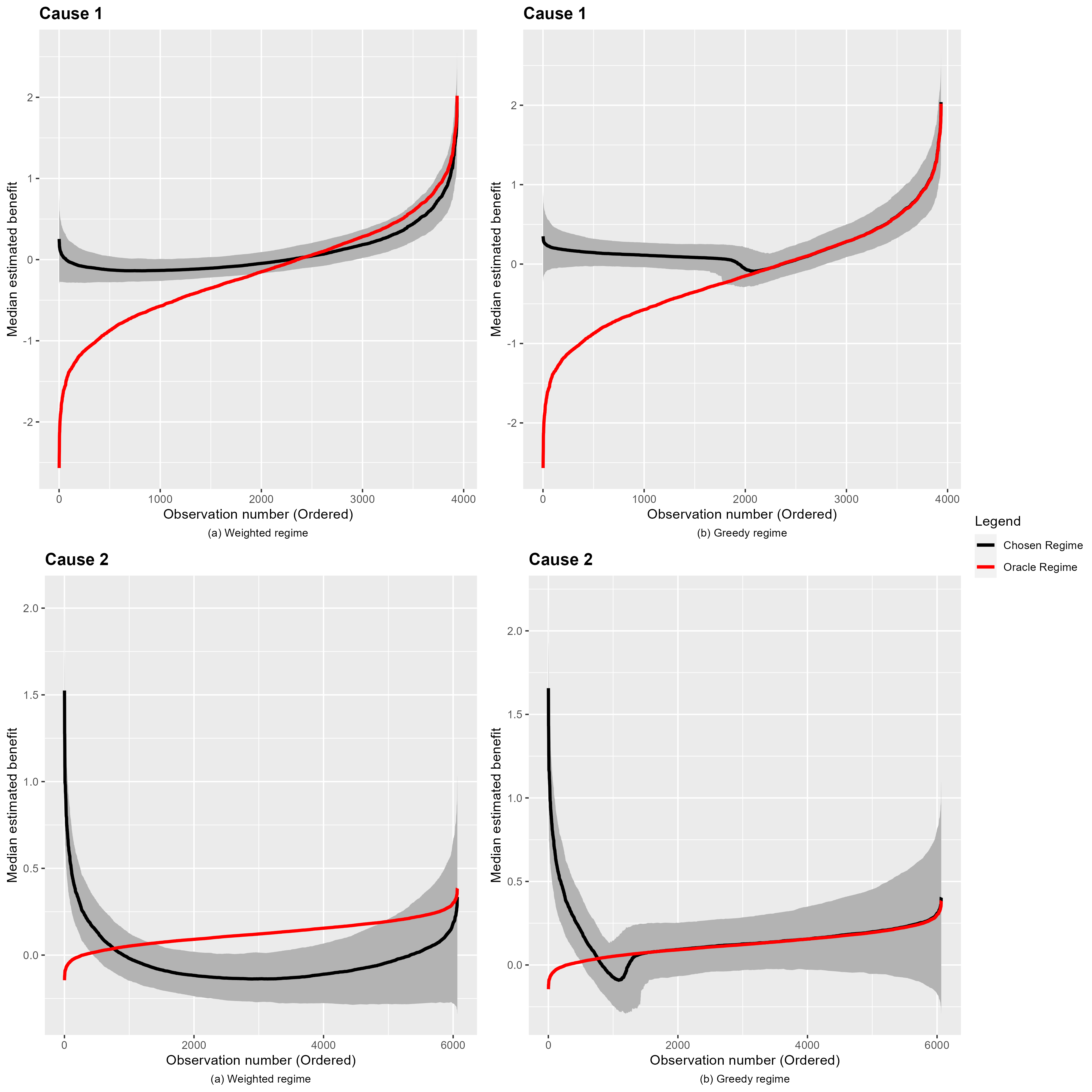}
    \caption*{Figure A10: Median estimated benefit for the weighted (left column) and greedy (right column) regimes along with the benefit of the oracle regime for setting 3, for causes $K=1$ (top row) and $K=2$ (bottom row). Benefit curves were evaluated on the test set, with subjects ordered by their benefit in increasing order. Estimates were computed over 1000 replicate datasets.}
   \label{fig:comparison_s2}
 \end{figure}

  \begin{figure}[H]
    \centering
    \includegraphics[scale = 0.30]{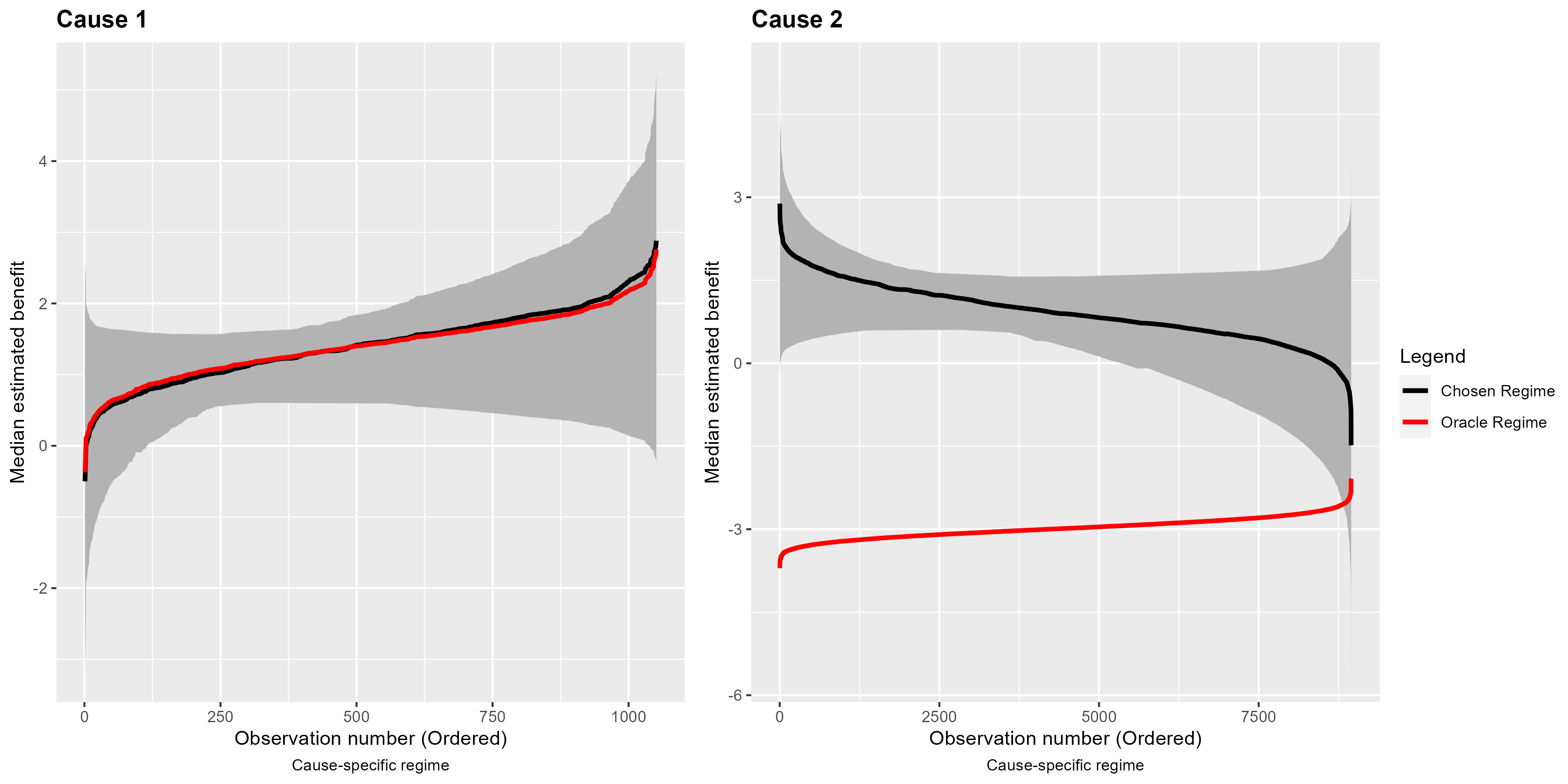}
    \caption*{Figure A11: Median estimated benefit for the cause-specific regime along with the benefit of the oracle regime for setting $10$, for causes $K=1$ (left) and $K=2$ (right). Benefit curves were evaluated on the test set, with subjects ordered by their benefit in increasing order. Estimates were computed over 1000 replicate datasets.}
   \label{fig:cause_specific}
 \end{figure}

 \begin{figure}[H]
    \centering
    \includegraphics[scale = 0.3]{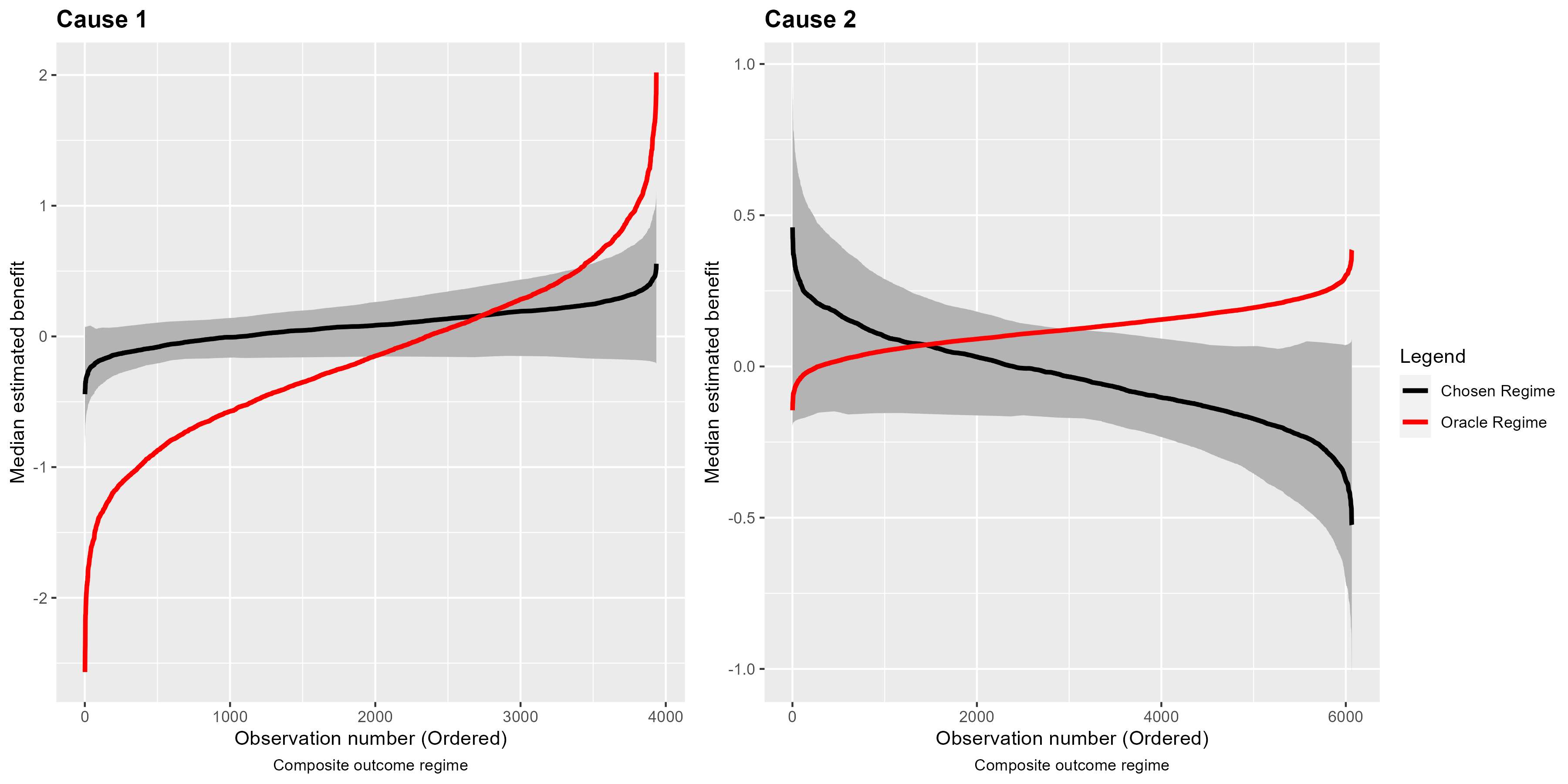}
    \caption*{Figure A12: Median estimated benefit for the composite outcome regime along with the benefit of the oracle regime for setting $4$, for causes $K=1$ (left) and $K=2$ (right). Benefit curves were evaluated on the test set, with subjects ordered by their benefit in increasing order. Estimates were computed over 1000 replicate datasets.}
   \label{fig:composite}
 \end{figure}

\subsection*{Additional simulations}
\label{subsec:other_sims}

This set of simulations explores the sensitivity of the proposed estimation procedure to variations in the data generating process and other modelling choices. For settings $4$-$9$, we use the following parameter values: 
\begin{align*}
    (\psi_{11}, \psi_{12}) &= (0.6, -0.6), \quad (\psi_{21}, \psi_{22}) = (-0.6, -0.6),\quad
    \tau^2 + \sigma^2 = 1,
\end{align*}
and consider the following settings:
{\allowdisplaybreaks
\begin{align*}
    &\text{Setting } 4: \text{As in section $3.1$, with } \delta_0 = 1.73 \text{ and } \text{ICC} = 0.5,\\
    &\text{Setting } 5.1: \text{ICC} = 0.9,\\
    &\text{Setting } 5.2: \text{ICC} = 0.1,\\
    &\text{Setting } 6: U_i\sim \text{Gamma}(\tau^2, 1) - \tau^2,\\
    &\text{Setting } 7: \delta_0 = 0, \text{corresponding to }50\% \text{ censoring},\\
    &\text{Setting } 8: \text{Independence working correlation matrix, i.e., } \mathbf{V}_i = \text{diag} (\lambda)_{j = 1, \dots, r_i}.
\end{align*}
}
For the next three simulation settings, we use the default specification of setting $4$ and consider clustering in the treatment assignment, with the same cluster indices as for the outcome model. To induce dependence in treatment values, we add a random intercept to the logistic regression model for $A_{ij}$. For cluster $i = 1,\dots, r$ and subject $j = 1,\dots, r_i$, we generate values of the treatment according to
\begin{align*}
    E_i &\sim \mathcal{N}(0, \xi^2), \quad 
    A_{ij} \sim \text{Bernoulli}\Bigl(\text{expit}(0.5 + X_{ij1} + X_{ij2} + E_i)\Bigr).
\end{align*}
The following values for the variance of the random intercept were considered, each corresponding to a different degree of within-cluster dependence of treatments:
\begin{align*}
    &\text{Setting } 9.1, \;  9.2, \;  9.3: \quad \xi^2 = ( 0.01, 0.25, 1).
\end{align*}
Standard errors, biases, and metrics for the derived ITRs are found in table S1. As expected, the standard errors for the blip estimators were larger when the degree of censoring was more substantial (setting 7); SEs were also larger when the working correlation structure was misspecified (setting 8). For fixed overall variance $\tau^2 + \sigma^2$, SEs were larger for smaller values of ICC, corresponding to larger within-cluster variability (settings 5.1 and 5.2). Standard errors were generally unaffected by the specification of the random effect distribution (setting 6) and by clustering within treatment (settings 9.1, 9.2 and 9.3). As for the derived ITRs, both greedy and weighted rules performed very similarly across all settings, with reasonable POTs and values close to those of the oracle regime. Benefit plots are presented below.

\begin{table}[H]
    \centering
     \caption*{Table S1: Root-$n$ adjusted bias and standard errors for blip parameter estimators, as well as ITR metrics for settings $4$-$10$. Estimates were computed using 1000 replicate datasets. Regimes are identified with $d_w$, $d_g$, $d_{cs}$, $d_o$, and $d_u$ representing the weighted, greedy, cause-specific, oracle and uniform regimes, respectively.\\}
    \begin{tabular}{|l|c|c|c|c|c|c|c|c|c|c|c|}
    \hline
        \multirow{2}{*}{} & \multirow{2}{*}{$\bm{\psi}$}& \multicolumn{10}{|c|}{Setting}\\
        & & 4 & 5.1 & 5.2 & 6 & 7 & 8 & 9.1 & 9.2 & 9.3 & 10\\
        \hline
         $\sqrt{n}\; \times$ Bias & $\psi_{11}$& -0.10& -0.16 &-0.07 &0.09 &0.01 &-0.13 &0.25 &0.31 &0.11 &-0.28 \\
         & $\psi_{12}$ &-0.14 &-0.07&-0.19 &-0.07 &0.05 &-0.26 &-0.00 &0.05 &-0.13&0.67\\
         & $\psi_{21}$ &0.08 & 0.07&0.05 &0.03 &-0.28 &0.01 &0.02 &0.08 &-0.07&0.03\\
         & $\psi_{22}$ &0.11 &0.05 &0.16 &0.25 &-0.17 &0.14 &0.07 &-0.14 &0.02&0.22 \\
         \hline
         $\sqrt{n}\; \times$ SE & $\psi_{11}$&4.65 &3.88 &5.22 &4.90 &7.04 &5.20 &4.90 &4.68 &4.89&8.14\\
         & $\psi_{12}$ &4.77 &4.12 &5.24 &4.89 &7.16 &5.16 &4.76 &4.54 &4.88&7.49\\
         & $\psi_{21}$ &3.59 &2.96 &3.99 &3.58 &4.96 &3.98 &3.32 &3.59 &3.32&1.95 \\
         & $\psi_{22}$ &4.35 &3.86 &4.66 &4.18 &6.46 &4.81 &4.29 &4.10 &4.41&2.38 \\
         \hline
          \multicolumn{2}{|c|}{POT($\hat{d}_{\text{w}}$)}& 0.74& 0.74 &0.74 &0.74 &0.74 &0.74 &0.74 &0.74 &0.74&0.90\\
          \multicolumn{2}{|c|}{POT($\hat{d}_{\text{g}}$)}&0.75 &0.75 &0.75 &0.75 &0.75 &0.75 &0.76 &0.76 &0.76&0.90\\
         \multicolumn{2}{|c|}{POT($\hat{d}_\text{cs}$)}&- &- &- &- &- &- &- &- &-&0.19\\
         \hline
          \multicolumn{2}{|c|}{V($\hat{d}_{\text{w}}$)}&1.82 &1.84 &1.78 &1.82 &1.81 &1.82 &1.74 &1.73 &1.74&1.87\\
          \multicolumn{2}{|c|}{V($\hat{d}_{\text{g}}$)}&1.82 &1.82 &1.78 &1.79 &1.79 &1.81 &1.73 &1.73 &1.73&1.87\\
          \multicolumn{2}{|c|}{V($\hat{d}_{\text{cs}}$)}&- &- &- &- &- &- &- &- &-& -0.65\\
         \hline
          \multicolumn{2}{|c|}{V($d_{\text{o}}$)}&1.92 &1.93 &1.89 &1.92 &1.91 &1.92 &1.83 &1.83 &1.83&2.02\\
          \multicolumn{2}{|c|}{V($d_{\text{u}}$)}&1.48 &1.50 &1.45 &1.48 &1.48 &1.48 &1.40 &1.40 &1.40&0.61\\
         \hline
    \end{tabular}
   
\end{table}

  \begin{figure}[h]
    \centering
    \includegraphics[scale = 0.30]{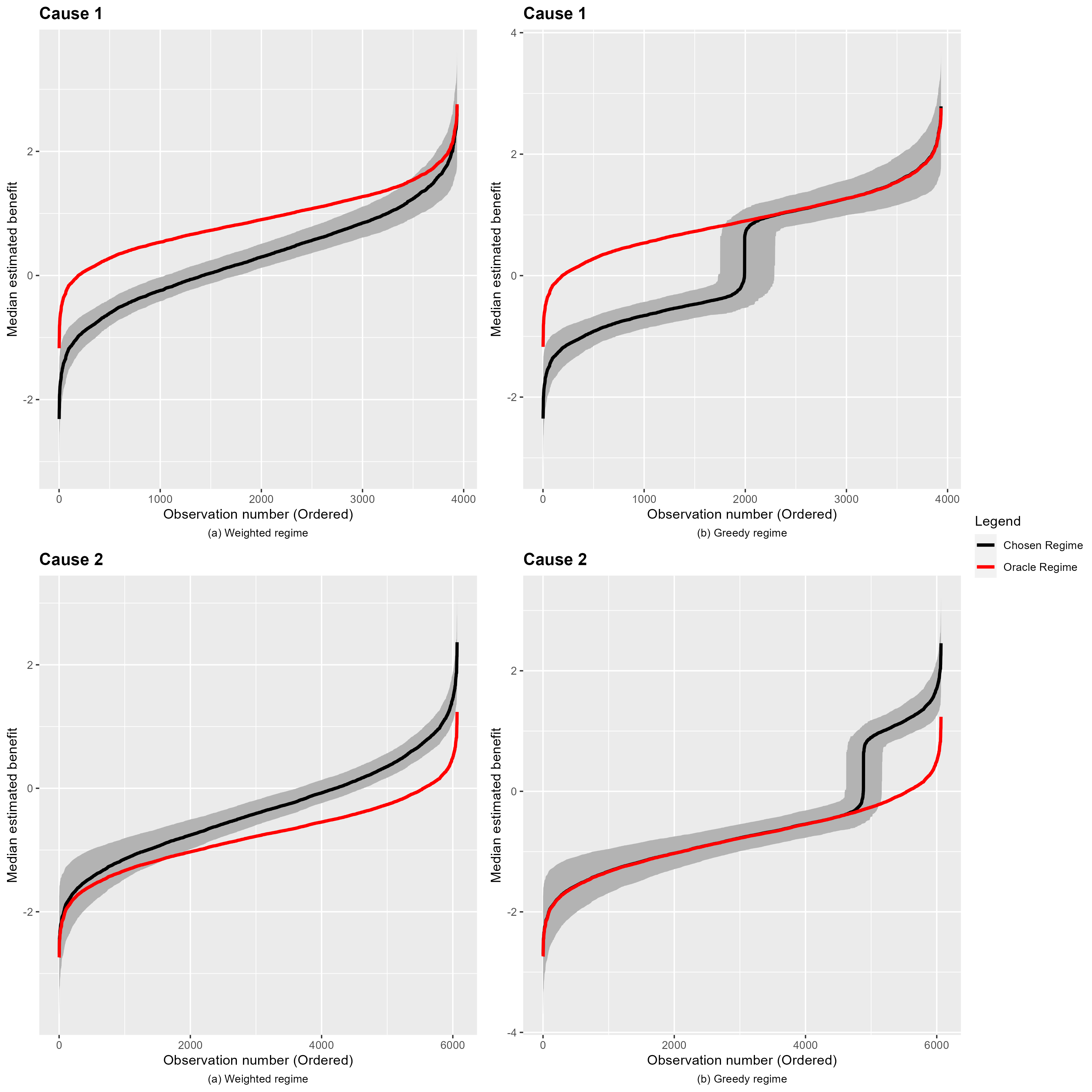}
    \caption*{Figure A13: Median estimated benefit for the weighted (left column) and greedy (right column) regimes along with the benefit of the oracle regime for setting 4, for causes $K=1$ (top row) and $K=2$ (bottom row). Benefit curves were evaluated on the test set, with subjects ordered by their benefit in increasing order. Estimates were computed over 1000 replicate datasets.}
 \end{figure}

 \begin{figure}[H]
    \centering
    \includegraphics[scale = 0.30]{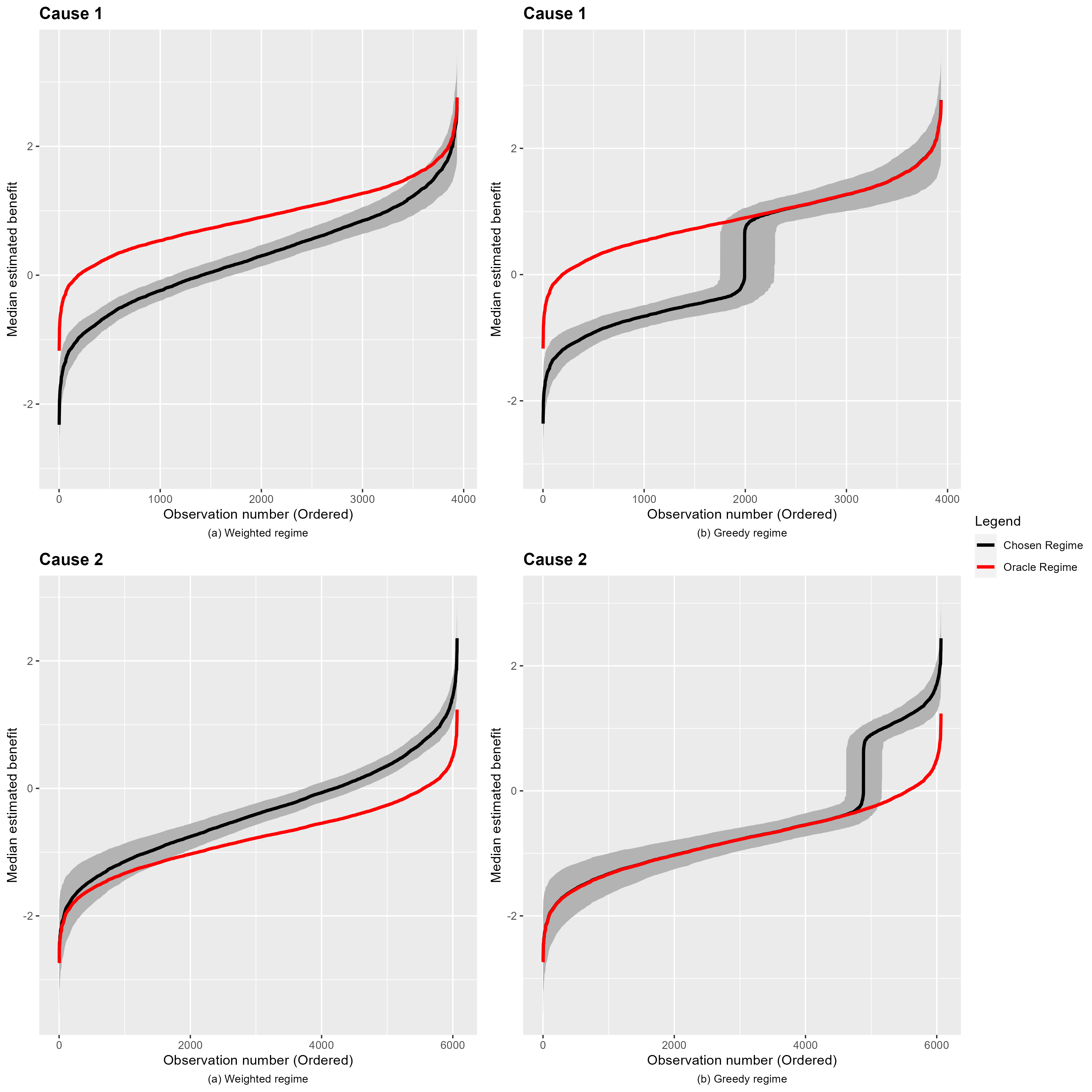}
    \caption*{Figure A14: Median estimated benefit for the weighted (left column) and greedy (right column) regimes along with the benefit of the oracle regime for setting 5.1, for causes $K=1$ (top row) and $K=2$ (bottom row). Benefit curves were evaluated on the test set, with subjects ordered by their benefit in increasing order. Estimates were computed over 1000 replicate datasets.}
 \end{figure}

 \begin{figure}[H]
    \centering
    \includegraphics[scale = 0.30]{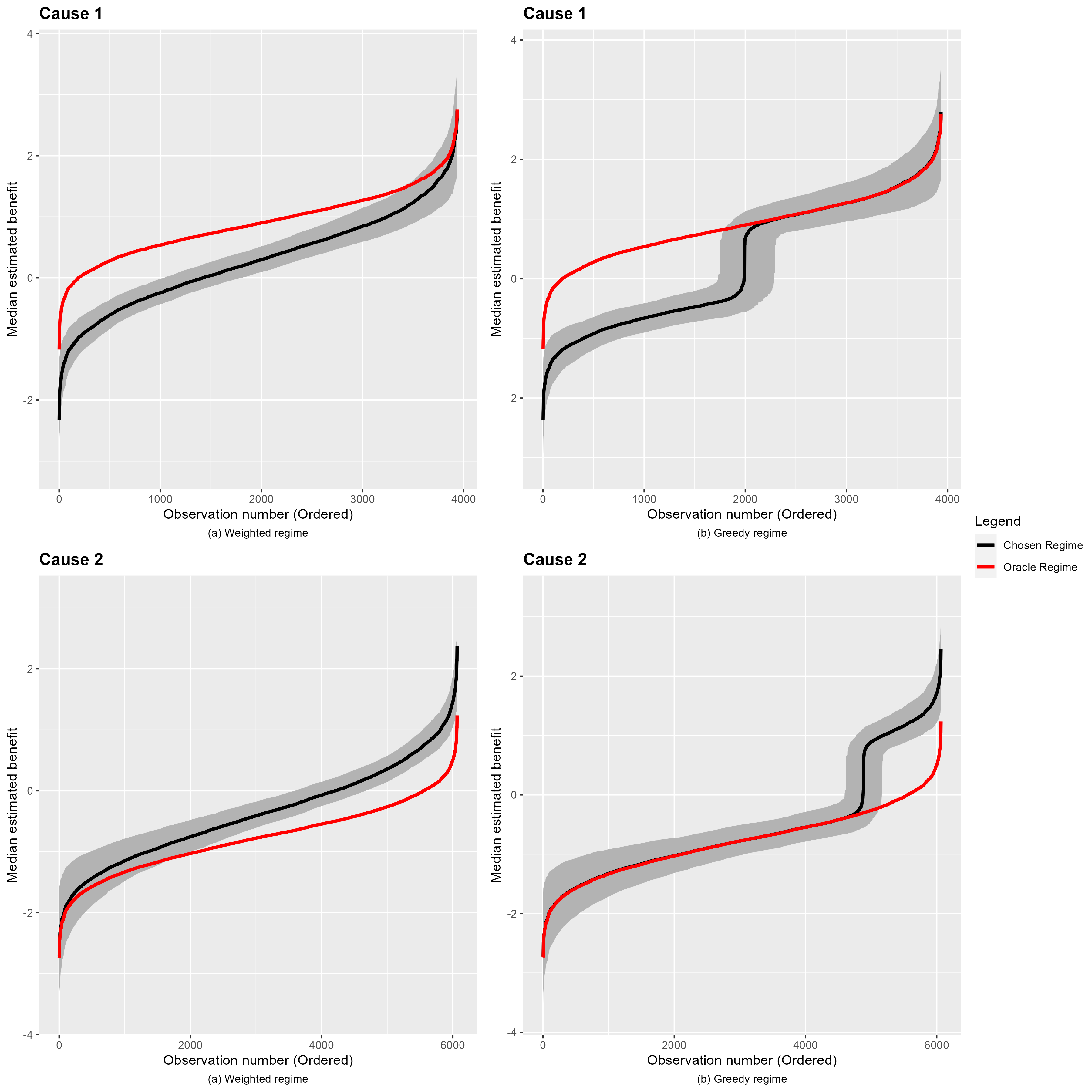}
    \caption*{Figure A15: Median estimated benefit for the weighted (left column) and greedy (right column) regimes along with the benefit of the oracle regime for setting 5.2, for causes $K=1$ (top row) and $K=2$ (bottom row). Benefit curves were evaluated on the test set, with subjects ordered by their benefit in increasing order. Estimates were computed over 1000 replicate datasets.}
 \end{figure}

  \begin{figure}[H]
    \centering
    \includegraphics[scale = 0.30]{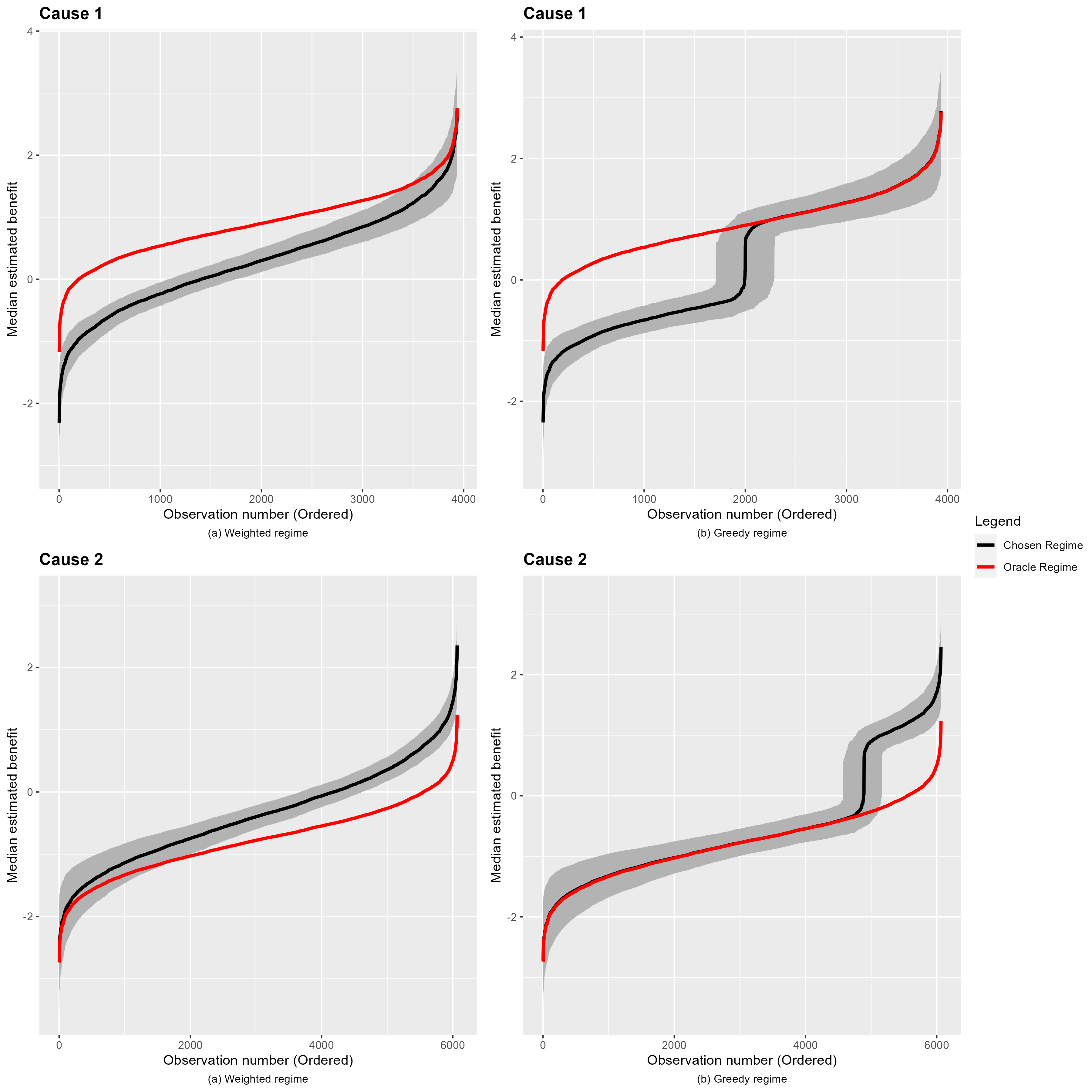}
    \caption*{Figure A16: Median estimated benefit for the weighted (left column) and greedy (right column) regimes along with the benefit of the oracle regime for setting 6, for causes $K=1$ (top row) and $K=2$ (bottom row). Benefit curves were evaluated on the test set, with subjects ordered by their benefit in increasing order. Estimates were computed over 1000 replicate datasets. }
 \end{figure}

 \begin{figure}[H]
    \centering
    \includegraphics[scale = 0.30]{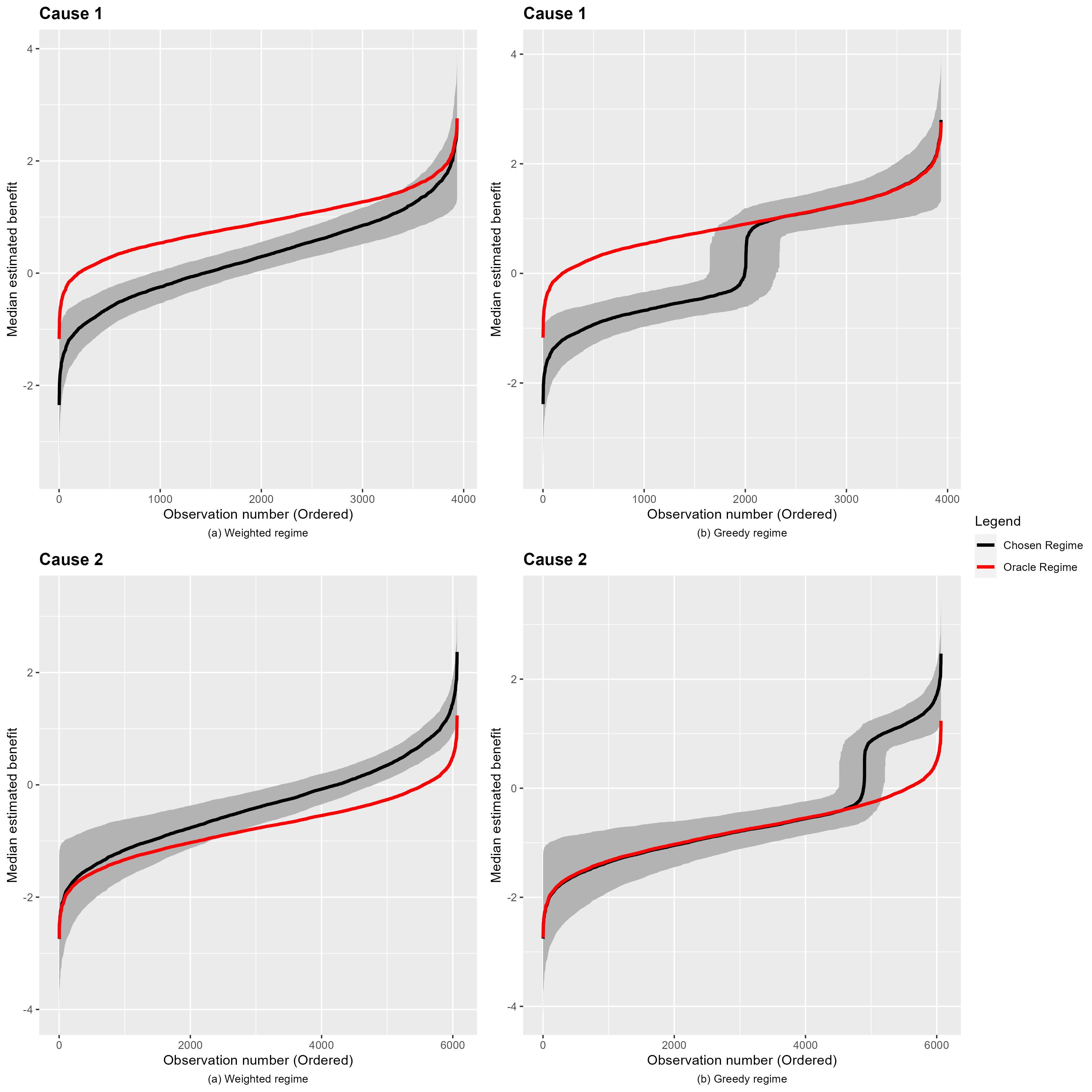}
    \caption*{Figure A17: Median estimated benefit for the weighted (left column) and greedy (right column) regimes along with the benefit of the oracle regime for setting 7, for causes $K=1$ (top row) and $K=2$ (bottom row). Benefit curves were evaluated on the test set, with subjects ordered by their benefit in increasing order. Estimates were computed over 1000 replicate datasets. }
 \end{figure}

  \begin{figure}[H]
    \centering
    \includegraphics[scale = 0.30]{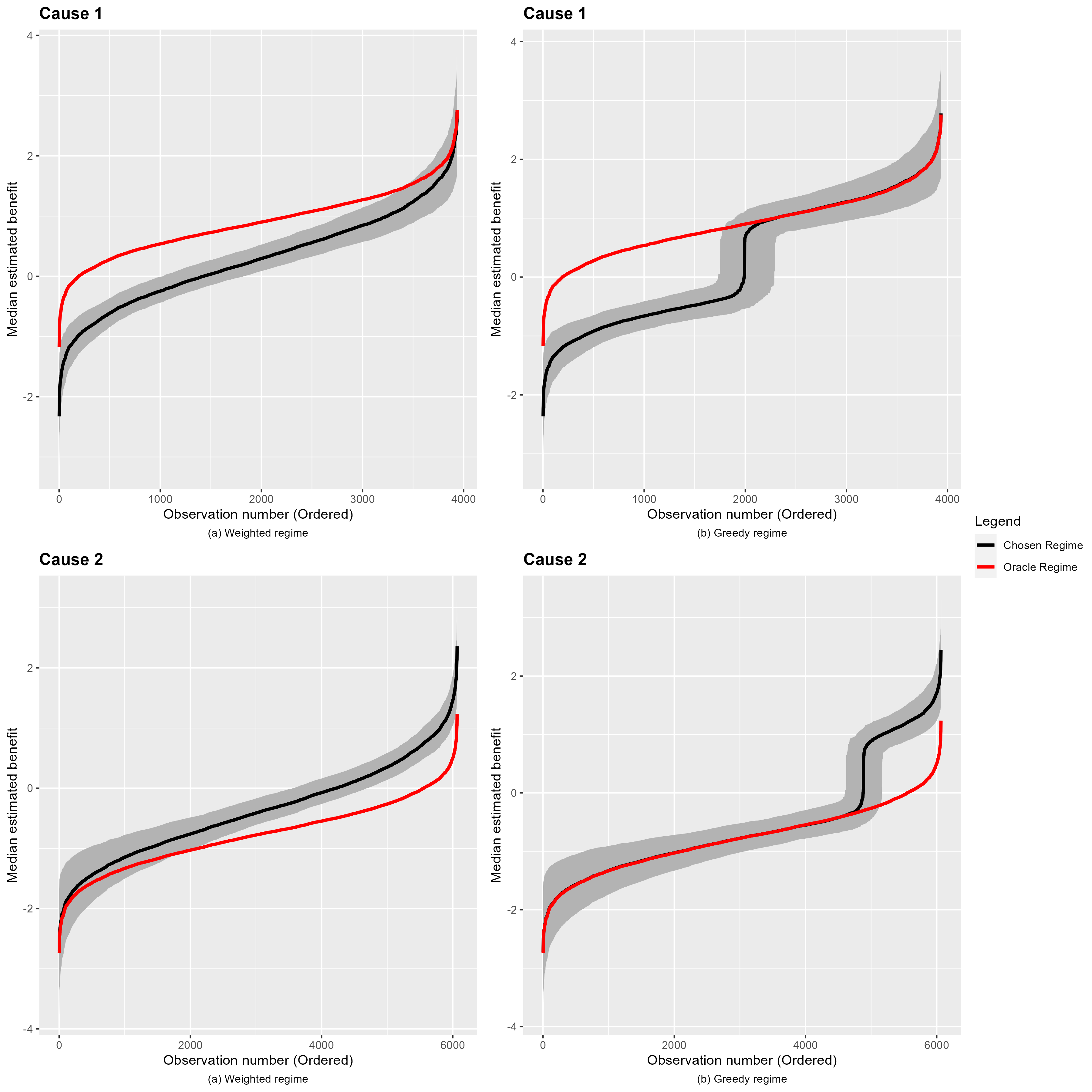}
    \caption*{Figure A18: Median estimated benefit for the weighted (left column) and greedy (right column) regimes along with the benefit of the oracle regime for setting 8, for causes $K=1$ (top row) and $K=2$ (bottom row). Benefit curves were evaluated on the test set, with subjects ordered by their benefit in increasing order. Estimates were computed over 1000 replicate datasets.}
 \end{figure}

 \begin{figure}[H]
    \centering
    \includegraphics[scale = 0.30]{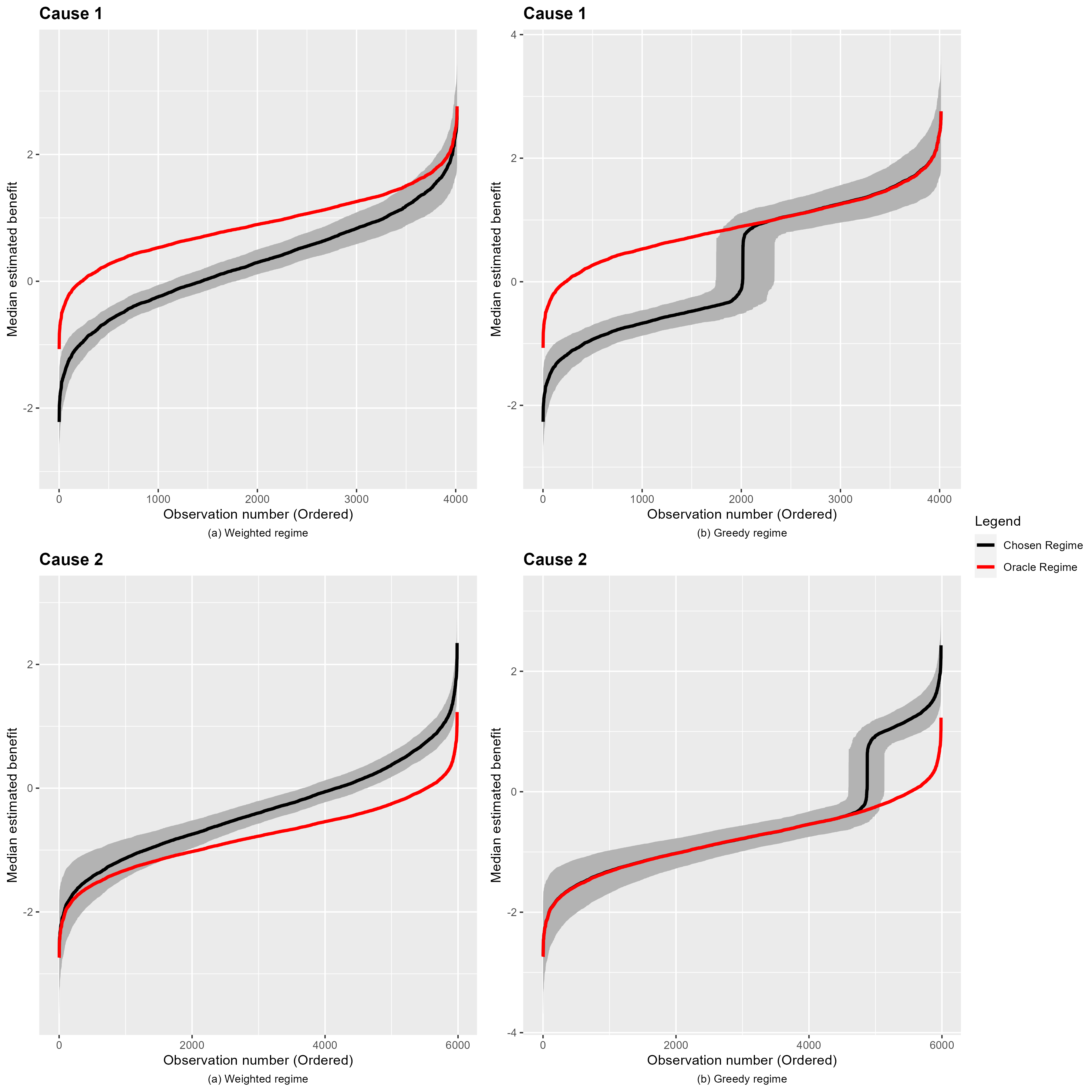}
    \caption*{Figure A19: Median estimated benefit for the weighted (left column) and greedy (right column) regimes along with the benefit of the oracle regime for setting 9.1, for causes $K=1$ (top row) and $K=2$ (bottom row). Benefit curves were evaluated on the test set, with subjects ordered by their benefit in increasing order. Estimates were computed over 1000 replicate datasets.}
 \end{figure}

 \begin{figure}[H]
    \centering
    \includegraphics[scale = 0.30]{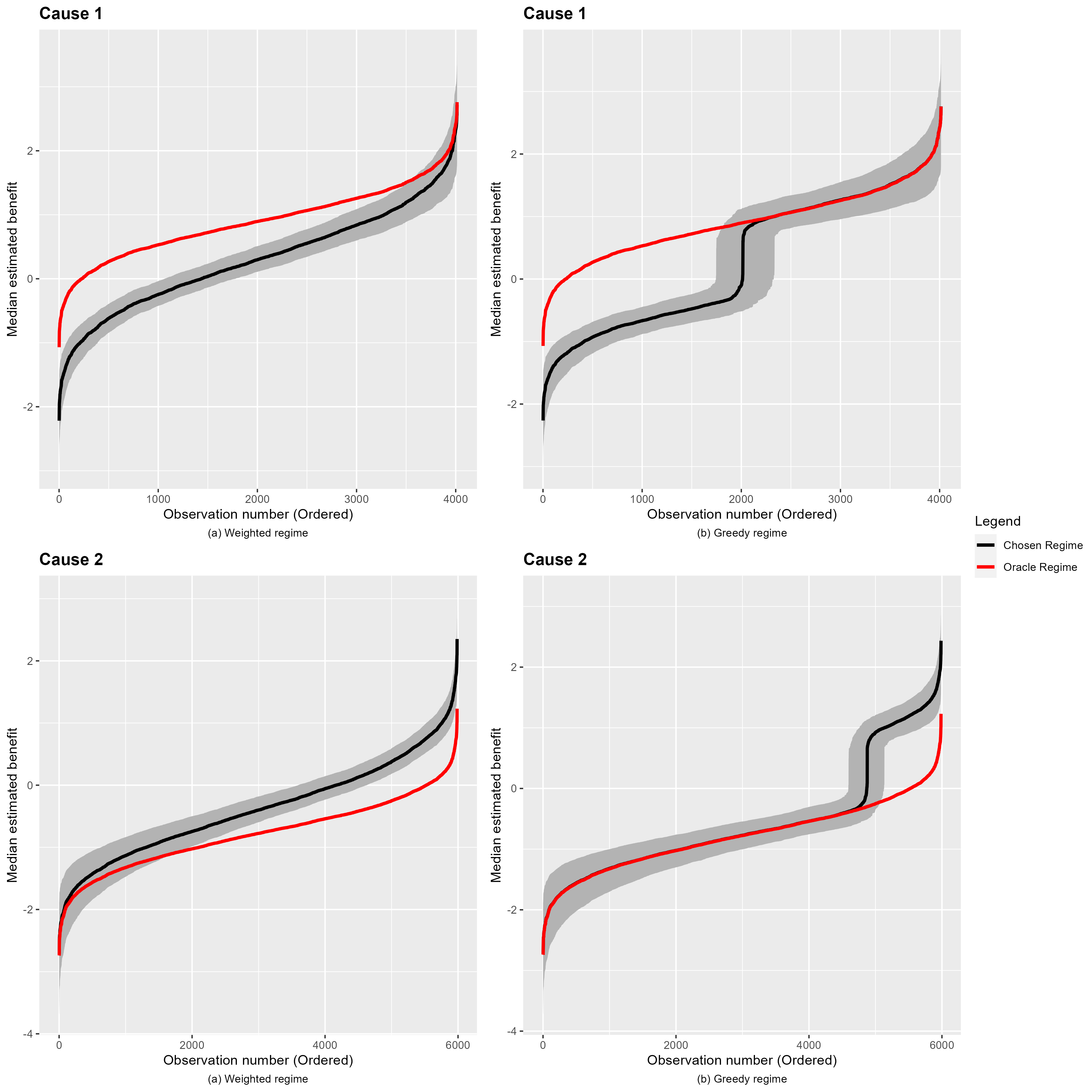}
    \caption*{Figure A20: Median estimated benefit for the weighted (left column) and greedy (right column) regimes along with the benefit of the oracle regime for setting 9.2, for causes $K=1$ (top row) and $K=2$ (bottom row). Benefit curves were evaluated on the test set, with subjects ordered by their benefit in increasing order. Estimates were computed over 1000 replicate datasets. }
 \end{figure}
 
 \begin{figure}[H]
    \centering
    \includegraphics[scale = 0.30]{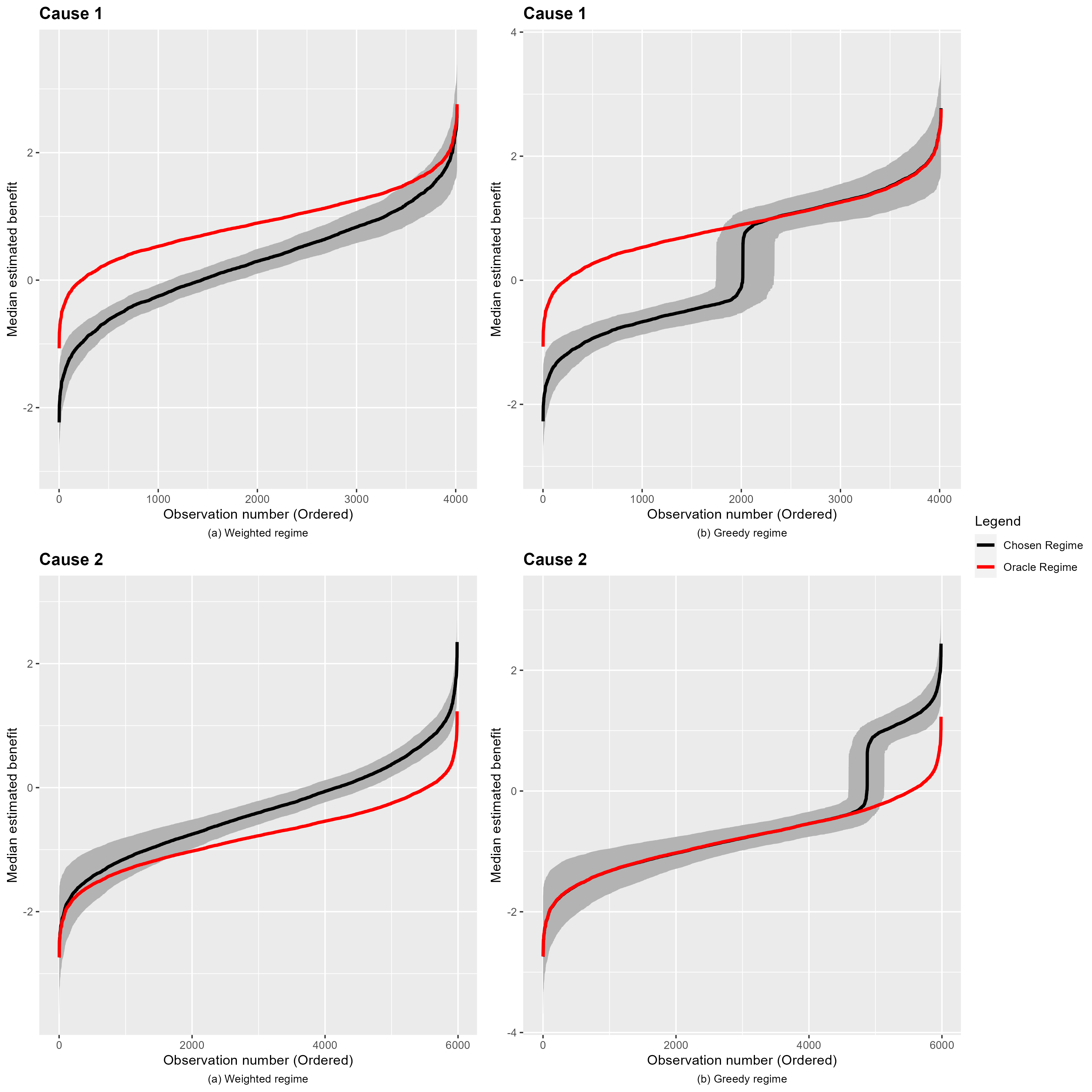}
    \caption*{Figure A21: Median estimated benefit for the weighted (left column) and greedy (right column) regimes along with the benefit of the oracle regime for setting 9.3, for causes $K=1$ (top row) and $K=2$ (bottom row). Benefit curves were evaluated on the test set, with subjects ordered by their benefit in increasing order. Estimates were computed over 1000 replicate datasets.}
 \end{figure}


\pagenumbering{gobble} 

 \section*{Appendix B}

 The following models were used for the data analysis: 
 \begin{align*}
     P(\text{DonHCV} = 1) = \text{expit}\Bigl(&\text{RecAge} + \text{RecGender} + \text{RecAge} + \text{RecEthnicity} + \text{RecDiabetes} +\text{RecHCV} + \\ &\text{RecImmunoGroup} + \log \text{CenterSize} \Bigr)\\
      P(\Delta= 1)= \text{expit}\Bigl(&\text{RecAge} + \text{RecGender} + \text{RecAge} + \text{RecEthnicity} + \text{RecDiabetes} +\text{RecHCV} + \\ &\text{RecImmunoGroup} +  \log \text{CenterSize} + \text{DonorHCV} + \text{DonorType} + \text{DonorAge} + \\&
      \text{DonorEthnicity} +  \text{DonorGender} +  \text{DonorCigUse} +  \text{DonorStroke} +  \\&
      \text{KidneyIschemicTime} + \text{OrganAllocationType}\Bigr)\\
      E[T| K=k] = &\text{RecAge} + \text{RecGender} + \text{RecAge} + \text{RecEthnicity} + \text{RecDiabetes} +\text{RecHCV} + \\ &\text{RecImmunoGroup} +  \log \text{CenterSize} + \text{DonorHCV} + \text{DonorType} + \text{DonorAge} + \\&
      \text{DonorEthnicity} +  \text{DonorGender} +  \text{DonorCigUse} +  \text{DonorStroke} +  \\&
      \text{KidneyIschemicTime} + \text{OrganAllocationType} + \\&
      \text{DonorHCV} \times \Bigl(\text{RecHCV} + \text{DonorType} + \text{RecAge}\Bigr),\quad k = 1,2\\
      P(K = 1) =\text{expit}\Bigl(&\text{RecAge} + \text{RecGender} + \text{RecAge} + \text{RecEthnicity} + \text{RecDiabetes} +\text{RecHCV} + \\ &\text{RecImmunoGroup} +  \log \text{CenterSize} + \text{DonorHCV} + \text{DonorType} + \text{DonorAge} + \\&
      \text{DonorEthnicity} +  \text{DonorGender} +  \text{DonorCigUse} +  \text{DonorStroke} +  \\&
      \text{KidneyIschemicTime} + \text{OrganAllocationType}\Bigr).
 \end{align*}
 Donor and recipient variables are prefixed by Don and Rec, respectively.

  \begin{table}[h]

  \caption*{Table S1: OPTN kidney transplant characteristics stratified by donor HCV status along with standardized mean differences (SMD).}
\centering
\begin{tabular}{lllll}
  \hline
  Characteristics & HCV- & HCV+ & Overall & SMD \\ 
  \hline
 & ($n$=300,161) & ($n$=11,313) & ($n$=311,474) &  \\ 
  Recipient gender &  &  &  & 0.249 \\ 
    \quad Female & 119857 (39.9\%) & 3190 (28.2\%) & 123047 (39.5\%) &  \\ 
    \quad Male & 180304 (60.1\%) & 8123 (71.8\%) & 188427 (60.5\%) &  \\ 
  Recipient ethnicity &  &  &  & 0.399 \\ 
   \quad Non-African American & 220256 (73.4\%) & 6179 (54.6\%) & 226435 (72.7\%) &  \\ 
    \quad African American & 79905 (26.6\%) & 5134 (45.4\%) & 85039 (27.3\%) &  \\ 
  Immuno-group &  &  &  & 0.099 \\ 
    \quad 1 & 186856 (62.3\%) & 7518 (66.5\%) & 194374 (62.4\%) &  \\ 
    \quad 2 & 63193 (21.1\%) & 2012 (17.8\%) & 65205 (20.9\%) &  \\ 
    \quad 3 & 9908 (3.3\%) & 410 (3.6\%) & 10318 (3.3\%) &  \\ 
   \quad  4 & 40204 (13.4\%) & 1373 (12.1\%) & 41577 (13.3\%) &  \\ 
  Recipient HCV status &  &  &  & 0.922 \\ 
    \quad HCV- & 290861 (96.9\%) & 7191 (63.6\%) & 298052 (95.7\%) &  \\ 
    \quad HCV+ & 9300 (3.1\%) & 4122 (36.4\%) & 13422 (4.3\%) &  \\ 
  Donor type &  &  &  & 0.726 \\ 
    \quad Deceased & 214903 (71.6\%) & 10925 (96.6\%) & 225828 (72.5\%) &  \\ 
    \quad Living & 85258 (28.4\%) & 388 (3.4\%) & 85646 (27.5\%) &  \\ 
  Recipient age (years) &  &  &  & 0.547 \\ 
    \quad Mean (SD) & 49.8 (15.6) & 57.1 (10.9) & 50.0 (15.5) &  \\ 
  Donor age (years) &  &  &  & 0.467 \\ 
    \quad 0-18 & 22956 (7.6\%) & 44 (0.4\%) & 23000 (7.4\%) &  \\ 
    \quad 19-49 & 197611 (65.8\%) & 9314 (82.3\%) & 206925 (66.4\%) &  \\ 
    \quad $>$ 49 & 79594 (26.5\%) & 1955 (17.3\%) & 81549 (26.2\%) &  \\ 
  Donor ethnicity &  &  &  & 0.176 \\ 
    \quad Non-African American & 261997 (87.3\%) & 10473 (92.6\%) & 272470 (87.5\%) &  \\ 
    \quad African American & 38164 (12.7\%) & 840 (7.4\%) & 39004 (12.5\%) &  \\ 
  Donor gender &  &  &  & 0.151 \\ 
   \quad  Female & 137046 (45.7\%) & 4324 (38.2\%) & 141370 (45.4\%) &  \\ 
    \quad Male & 163115 (54.3\%) & 6989 (61.8\%) & 170104 (54.6\%) &  \\ 
  Donor history of cigarette use &  &  &  & 0.338 \\ 
    \quad No & 232353 (77.4\%) & 7027 (62.1\%) & 239380 (76.9\%) &  \\ 
    \quad Yes & 67808 (22.6\%) & 4286 (37.9\%) & 72094 (23.1\%) &  \\ 
  Donor cerebrovascular/stroke &  &  &  & 0.113 \\ 
   \quad  No & 237035 (79.0\%) & 9431 (83.4\%) & 246466 (79.1\%) &  \\ 
   \quad  Yes & 63126 (21.0\%) & 1882 (16.6\%) & 65008 (20.9\%) &  \\ 
  Kidney cold ischemic time (hours) &  &  &  & 0.507 \\ 
   \quad  Mean (SD) & 13.6 (10.5) & 18.5 (8.49) & 13.8 (10.5) &  \\ 
  Organ allocation type &  &  &  & 0.835 \\ 
    \quad Local & 38210 (12.7\%) & 3585 (31.7\%) & 41795 (13.4\%) &  \\ 
    \quad Regional & 0 (0\%) & 0 (0\%) & 0 (0\%) &  \\ 
    \quad National & 236221 (78.7\%) & 4642 (41.0\%) & 240863 (77.3\%) &  \\ 
    \quad Foreign & 25730 (8.6\%) & 3086 (27.3\%) & 28816 (9.3\%) &  \\ 
  Log of center size &  &  &  & 0.094 \\ 
    \quad Mean (SD) & 7.47 (0.811) & 7.55 (0.718) & 7.48 (0.808) &  \\ 
  Follow-up time (days) &  &  &  & 0.614 \\ 
    \quad Median [IQR] & 1526 [695, 2900] & 739 [355, 1469] & 1480 [676, 2875] &  \\ 
  Event type &  &  &  & 0.154 \\ 
   \quad  Censored & 203410 (67.8\%) & 8374 (74.0\%) & 211784 (68.0\%) &  \\ 
    \quad Graft failure & 48227 (16.1\%) & 1281 (11.3\%) & 49508 (15.9\%) &  \\ 
    \quad Death with functioning graft & 48524 (16.2\%) & 1658 (14.7\%) & 50182 (16.1\%) &  \\ 
   \hline
\end{tabular}
\end{table}

 \begin{figure}[h]
    \centering
    \includegraphics[scale = 0.10]{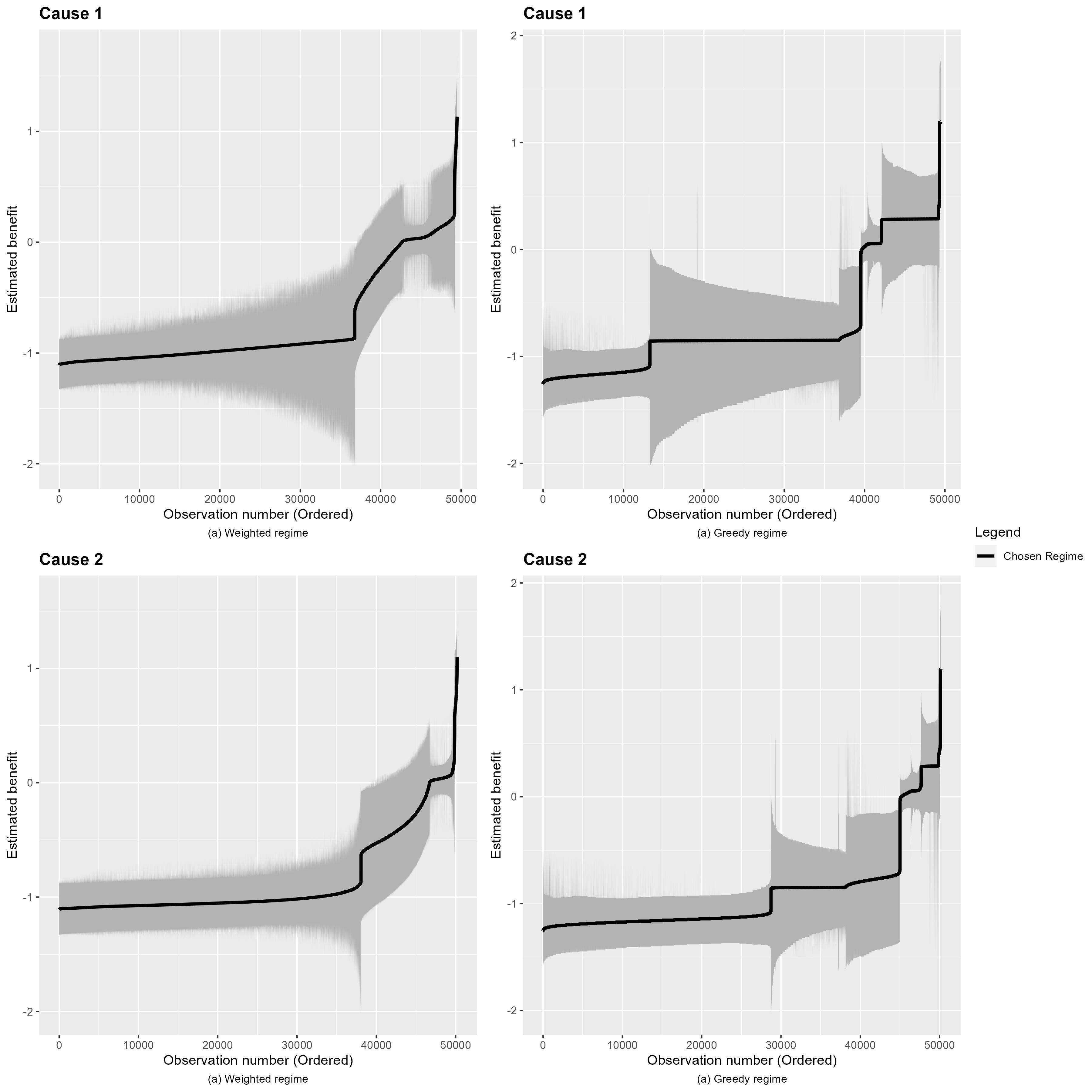}
    \caption*{Figure B1: Estimated benefit for the weighted (left column) and greedy (right column) regimes for causes $K=1$ (top row) and $K=2$ (bottom row), along with associated pointwise 95\% bootstrap confidence intervals. The analysis was conducted using the OPTN data on $n = 311,474$ individuals from $251$ centers, based on kidney transplantations carried out in the period from January 1, 2001 to December 31, 2022. Benefit curves were evaluated for all non-censored individuals, with subjects ordered by their increasing benefit. Bootstrap intervals were computed using $B = 1000$ replicate datasets.}
    
 \end{figure}

 \begin{table}[h]
    \centering
    \caption*{Figure B2: Blip parameter estimates and associated 95\% bootstrap confidence intervals (CI) for the cause-specific and composite outcome analyses using the OPTN data on $n = 311,774$ individuals from $251$ centers, based on kidney transplantations carried out in the period from January 1, 2001 to December 31, 2022. Nonparametric cluster bootstrap CIs were computed using $B=1000$ bootstrap replicates.\\}
    \begin{tabular}{|c|c|c|c|c|}
    \hline
     \multirow{2}{*}{Parameters}& \multicolumn{2}{|c|}{Cause-specific} & \multicolumn{2}{|c|}{Composite outcome}\\
     &  Estimate & 95\% CI & Estimate & 95\% CI\\
    \hline
    DonHCV & $-0.87$& $(-3.31, 6.26)$ & $-0.97$& $(-3.00, -0.53)$ \\
    DonHCVxRecHCV & $0.93$& $(-0.38, 1.80)$& $1.08$& $(0.96, 1.55)$\\
    DonHCVxDonType & $1.22$& $(-0.41, 1.89)$ & $0.82$& $(0.55, 1.23)$\\
    DonHCVxDonAge & $3.77\text{x} 10^{-5}$& $(-8.62\text{x}10^{-3}, 3.85\text{x} 10^{-3})$& $1.36\text{x}10^{-4}$& $(-7.02\text{x} 10^{-4}, 2.49\text{x}10^{-3})$\\
    \hline
    \end{tabular}
    \label{tab:data_analysis_estimates}
\end{table}

 \begin{figure}[h]
    \centering
    \includegraphics[scale = 0.10]{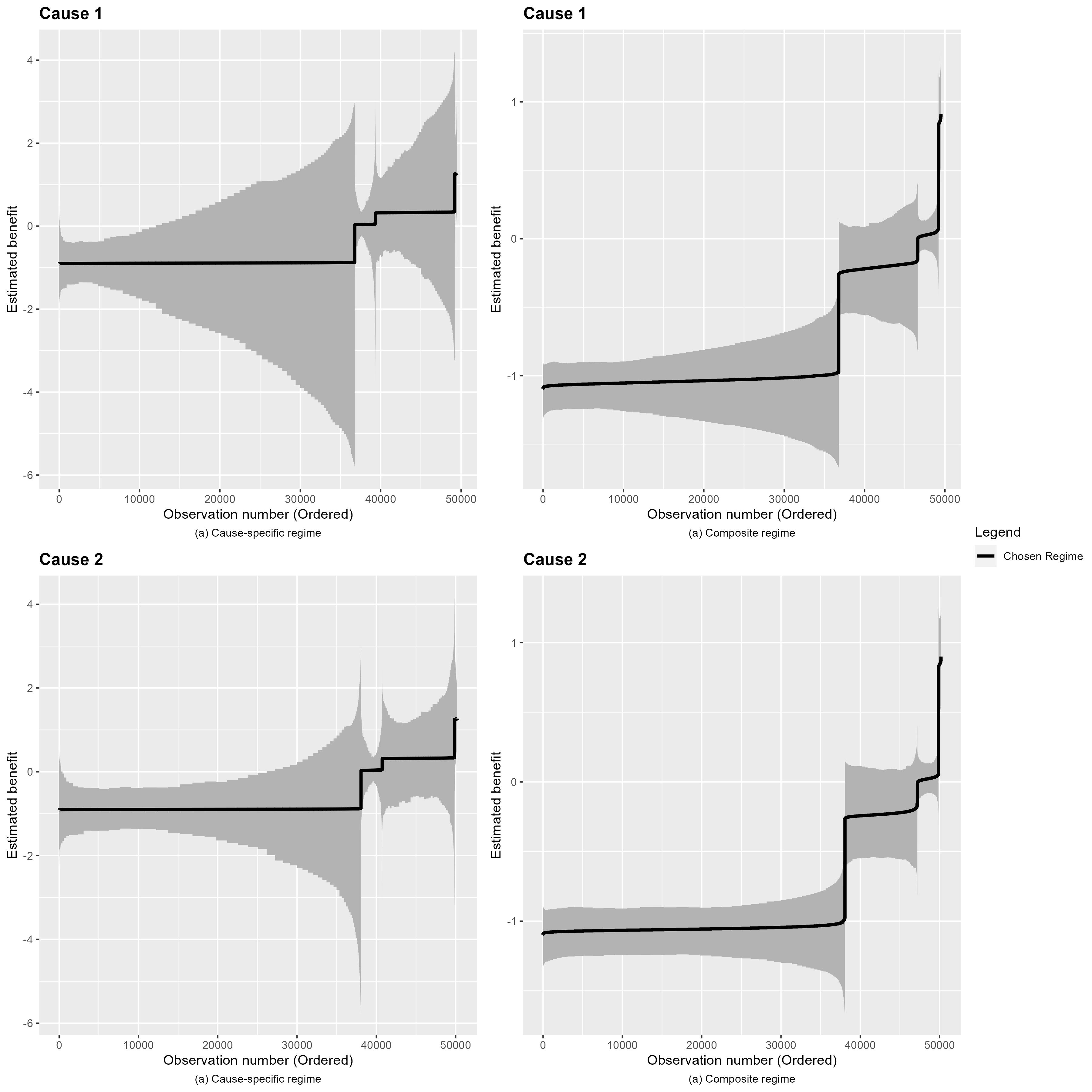}
    \caption*{Figure B3: Estimated benefit for the cause-specific (left column) and composite (right column) regimes for causes $K=1$ (top row) and $K=2$ (bottom row), along with associated pointwise 95\% bootstrap confidence intervals. The analysis was conducted using the OPTN data on $n = 311,474$ individuals from $251$ centers, based on kidney transplantations carried out in the period from January 1, 2001 to December 31, 2022. Benefit curves were evaluated for all non-censored individuals, with subjects ordered by their increasing benefit. Bootstrap intervals were computed using $B = 1000$ replicate datasets.}
    
 \end{figure}


\pagenumbering{gobble} 

\section*{Appendix C}

\subsection*{Proof of consistency of blip estimators}

Suppose we are estimating the blip parameters for cause $k$ and let $t_\text{max}$ denote the index of the last iteration of the AFT-GEE fitting procedure. The estimating equation used to estimate the blip parameters can be written in the following way:
\begin{align*}
                &\sum_{i=1}^r \mathbf{D}_i^\top [\mathbf{V}_i (\hat{\bm{\lambda}}^{(t_\text{max})})]^{-1}\; \mathbf{W_i}\Bigl(\log (\mathbf{T}_i) - \mathbf{X}_{i, \bm{\beta_k}}\bm{\beta_k} - \mathbf{A}\mathbf{X}_{i, \bm{\psi_k}}\bm{\psi_k}\Bigr),\\
                &= \sum_{i=1}^r \mathbf{D}_i^\top [\mathbf{V}_i (\hat{\bm{\lambda}}^{(t_\text{max})})]^{-1}\; \Bigl(\log (\mathbf{T}_i^w) - \mathbf{X}^w_{i, \bm{\beta_k}}\bm{\beta_k} - \mathbf{A}^w\mathbf{X}^w_{i, \bm{\psi_k}}\bm{\psi_k}\Bigr),
\end{align*}
which, by DWSurv theory, is equivalent to estimating $(\bm{\beta}_k, \bm{\psi}_K)$ by generalized least squares using the weighted dataset ($T^w, \bm{X}^w, A^w$), where the covariates $\bm{X}^w$ are independent of the treatment and censoring mechanism $(A^w, \Delta^w)$. By standard linear regression results, this is sufficient to account for possible omitted variable bias in $\bm{\psi}_k$ due to misspecification of the treatment-free model for any fixed variance-covariance matrices  $\mathbf{V}_i (\hat{\bm{\lambda}}^{(t_\text{max})}), \quad i = 1, \dots, r$. Thus, the procedure yields consistent estimators of the blip parameters $\bm{\psi}_k$ when the chosen weight function $w_k$ satisfies the DWSurv balancing property \citep{Simoneau2020}.

\subsection*{Estimation of ITR metrics}

For a given regime regime $d$, we estimate its POT and value via the following unbiased estimating equations in $p, \mu$:
\begin{align*}
    \frac{1}{n} \sum_{i=1}^n \frac{\Delta_i}{P(\Delta = 1 | \bm{X}_i, A_i)} \Bigl(\mathds{1}(d(\bm{X}_i) = d_{\text{o}}(\bm{X}_i, K_i)) - p \Bigr) &= 0,\\
    \frac{1}{n} \sum_{i=1}^n \frac{\Delta_i}{P(\Delta = 1 | \bm{X}_i, A_i)} \Bigl( \log T_i + (d(\bm{X}_i)-A_i) \bm{X}_{\psi_{K_i}} \psi_{K_i} - \mu \Bigr) &= 0.
\end{align*}
We now show that the two equations above are unbiased for the parameters of interest. For the POT of a given ITR $d$, we have that 
\begin{align*}
    &\mathbb{E}\Bigl[\frac{\Delta}{P(\Delta = 1 | \bm{X}, A)} \Bigl(\mathds{1}(d(\bm{X}) = d_{\text{o}}(\bm{X}, K)) - p \Bigr)\Bigr]\\
    &= \mathbb{E}\left[\mathbb{E}\Bigl[\frac{\Delta}{P(\Delta = 1 | \bm{X}, A)} \Bigl(\mathds{1}(d(\bm{X}) = d_{\text{o}}(\bm{X}, K)) - p \Bigr)\Bigl| \bm{X}, A, K\Bigr]\right]\\
    &=  \mathbb{E}\left[\Bigl(\mathds{1}(d(\bm{X}) = d_{\text{o}}(\bm{X}, K)) - p \Bigr) \mathbb{E}\Bigl[\frac{\Delta}{P(\Delta = 1 | \bm{X}, A)} \Bigl| \bm{X}, A, K\Bigr]\right]\\
    &= \mathbb{E}\left[\Bigl(\mathds{1}(d(\bm{X}) = d_{\text{o}}(\bm{X}, K)) - p \Bigr) \frac{P(\Delta = 1 | \bm{X}, A, K)}{P(\Delta = 1 | \bm{X}, A)}\right]\\
    &=\mathbb{E}\left[\Bigl(\mathds{1}(d(\bm{X}) = d_{\text{o}}(\bm{X}, K)) - p \Bigr) \frac{P(\Delta = 1 | \bm{X}, A)}{P(\Delta = 1 | \bm{X}, A)}\right] &\text{by assumption }5\\
    &=\mathbb{E}\left[\mathds{1}(d(\bm{X}) = d_{\text{o}}(\bm{X}, K))\right] - p\\
    &= 0,
\end{align*}
for $p = \text{POT}(d)$. \\

Similarly, for the value of a regime $d$,
\begin{align*}
   & \mathbb{E} \left[\frac{\Delta}{P(\Delta = 1 | \bm{X}, A)} \Bigl( \log T + (d(\bm{X})-A)  \bm{X}_{\psi_{K}} \bm{\psi}_{K} - \mu \Bigr)\right]\\
   &=\mathbb{E}\left[ \mathbb{E} \left[\frac{\Delta}{P(\Delta = 1 | \bm{X}, A)} \Bigl( \log T + (d(\bm{X})-A)  \bm{X}_{\psi_{K}} \bm{\psi}_{K} - \mu \Bigr) \Bigl| \bm{X},A,K, \Delta\right] \right]\\
    &= \mathbb{E}\left[ \frac{\Delta}{P(\Delta = 1 | \bm{X}, A)}  \mathbb{E} \left[\Bigl( \log T + (d(\bm{X})-A)  \bm{X}_{\psi_{K}} \bm{\psi}_{K} - \mu \Bigr) \Bigl| \bm{X},A,K, \Delta\right] \right].\\
\end{align*}
Then, we have the following identity:
\begin{align*}
    &\mathbb{E} \left[\Bigl( \log T + (d(\bm{X})-A)  \bm{X}_{\psi_{K}}\bm{\psi}_{K} - \mu \Bigr) \Bigl| \bm{X},A,K, \Delta\right]\\
    &=   \mathbb{E} \left[ \log T |\bm{X}, A, K, \Delta\right] + (d(\bm{X})-A)  \bm{X}_{\psi_{K}} \bm{\psi}_{K} - \mu\\
    &= \mathbb{E} \left[ \log T(A,K) |\bm{X}, A, K, \Delta\right] + (d(\bm{X})-A)  \bm{X}_{\psi_{K}} \bm{\psi}_{K} - \mu&\text{by assumptions }2, 3\\
    &= \mathbb{E} \left[ \log T(A,K) |\bm{X}, A, K\right] + (d(\bm{X})-A)  \bm{X}_{\psi_{K}} \bm{\psi}_{K} - \mu&\text{by assumption }4\\
    &= \mathbb{E} \left[ \log T |\bm{X}, A, K\right] + (d(\bm{X})-A)  \bm{X}_{\psi_{K}} \bm{\psi}_{K} - \mu\\
    &= \bm{X}_{\beta_K} \bm{\beta}_K + A \bm{X} \bm{\psi}_K + (d(\bm{X})-A)  \bm{X}_{\psi_{K}} \bm{\psi}_{K} - \mu\\
    &= \bm{X}_{\beta_K} \bm{\beta}_K +d(\bm{X})  \bm{X}_{\psi_{K}} \bm{\psi}_{K} - \mu\\
    &= \mathbb{E} \left[\log T \Bigl| \bm{X}, A =d, K\right] - \mu\\
    &=  \mathbb{E}\left[\log T(d,K)|\bm{X}\right] - \mu.
\end{align*}
Substituting this component back into the original equation gives:
\begin{align*}
    &\mathbb{E}\left[ \frac{\Delta}{P(\Delta = 1 | \bm{X}, A)}  \mathbb{E} \left[\Bigl( \log T + (d(\bm{X})-A)  \bm{X}_{\psi_{K}} \bm{\psi}_{K} - \mu \Bigr) \Bigl| \bm{X},A,K, \Delta\right] \right]\\
    &=   \mathbb{E}\left[ \frac{\Delta}{P(\Delta = 1 | \bm{X}, A)}  \left(\mathbb{E} \left[\log T(d,K)|\bm{X}\right]- \mu\right) \right]\\
    &= \mathbb{E} \left[\mathbb{E}\left[ \frac{\Delta}{P(\Delta = 1 | \bm{X}, A)} \left(\mathbb{E} \left[\log T(d,K)|\bm{X}\right]- \mu\right)\Bigl|\bm{X}, A\right]\right]\\
    &= \mathbb{E} \left[\left(\mathbb{E} \left[\log T(d,K)|\bm{X}\right]- \mu\right) \mathbb{E}\left[ \frac{\Delta}{P(\Delta = 1 | \bm{X}, A)} \Bigl|\bm{X}, A\right]\right]\\
    &=  \mathbb{E} \left[\mathbb{E} \left[\log T(d,K)|\bm{X}\right]\right] - \mu\\
    &= 0,
\end{align*}
for $\mu = \mathbb{E} [\log T(d,K)]$.

\subsection*{Extension to treatment-dependent causes}

We can derive alternative definitions of the optimal ITR if we replace assumptions $3$ and $5$ of section $2.1$ by the following weaker assumption:
\begin{enumerate}
    \item[*3.] Sequential ignorability \citep{imai2010}: for all $a \in \mathcal{A}$ and all $k = 1, \dots, \kappa$, 
\begin{align*}
    T(a,k) &\perp A | \bm{X},\\ 
    T(a,k) &\perp K | \bm{X}, A.
\end{align*}
\end{enumerate}
Under this assumption, the mediation formula of \cite{pearl2012} yields the following definition for the weighted ITR:
\begin{align*}
    d^{\text{opt}}_{\text{w}} &= \arg \max_{d} \mathbb{E}\Bigl[f(T(d,K))\Bigr] \\
    &= \arg \max_{d} \mathbb{E}\Bigl[\mathbb{E}[f(T(d,K))| \bm{X}] \Bigr]\\
    &=  \arg \max_{d} \mathbb{E}\left[ \sum_{k=1}^\kappa P\left(K = k| \bm{X}, A = d(\bm{X})\right) \mathbb{E}[f(T)| \bm{X}, A = d(\bm{X}), K = k]  \right].
\end{align*}   
Letting $\varphi_k(\bm{x}, a) = P(K = k| \bm{X} = \bm{x}, A = a)$ and $Q_k(\bm{x}, a) = \mathbb{E}[f(T)| \bm{X} = \bm{x}, A = a, K = k]$, we have the equivalent form:
\begin{align*}
     d^{\text{opt}}_{\text{w}} &=  \arg \max_{d} \mathbb{E}\left[ \sum_{k=1}^\kappa \varphi_k(\bm{X}, d(\bm{X})) Q_k(\bm{X}, d(\bm{X}))   \right],
\end{align*}
which gives rise to the explicit rule
\begin{align*}
     d_{\text{w}}(\bm{x}) &=  \arg \max_a \sum_{k=1}^\kappa \varphi_k(\bm{x}, a) Q_k(\bm{x}, a), \quad \bm{x} \in \mathcal{X}.
\end{align*}

\bibliographystyle{abbrvnat}
\bibliography{appendixC}